\documentclass[useAMS,usenatbib]{mn2e}  

\usepackage[]{graphicx}
\usepackage{latexsym}
\usepackage{amssymb}
\usepackage{float}
\usepackage{ifthen}
\usepackage{url}
\usepackage[utf8]{inputenc}
\usepackage{footmisc}
\usepackage{times}

\newcommand {\ha} {H$\alpha$}
\newcommand {\hb} {H$\beta$}
\newcommand {\hi} {H{\scriptsize I}}
\newcommand {\hii} {H{\scriptsize II}}
\newcommand {\hmol} {$\mathrm{H_2}$}
\newcommand {\msol}{M$_{\sun}$}
\newcommand {\oiii} {[O{\scriptsize III}]}

\newcommand{\Min}{${}^{\prime}$}
\newcommand{\Sec}{${}^{\prime\prime}$}
\newcommand{\Deg}{${}^{\circ}$}
\newcommand {\kms} {\,km\,s$^{-1}$}
\newcommand{\FM} {\texttt{\textsc{FaNTOmM}}}

\title[\ha\ Kinematics of the SINGS Galaxies]
  {\ha\ Kinematics of the SINGS Nearby Galaxies Survey \--- II}
\author[I. Dicaire et al.]
  {I.~Dicaire$^1$ \thanks{ Based on observations made with the ESO 3.60m telescope at La Silla Observatories under programme ID 076.B-0859 and on observations obtained at the Canada--France--Hawaii Telescope (CFHT) which is operated by the National Research Council of Canada, the Institut National des Sciences de l'Univers of the Centre National de la Recherche Scientifique of France,  and the University of Hawaii} \thanks{E--mail: isabelle@astro.umontreal.ca},
  C.~Carignan$^{1,7}$, P.~Amram$^{2,7}$,
  O.~Hernandez$^{1,7}$, L.~Chemin$^{3}$,
\newauthor
  O.~Daigle$^{1,2,7}$, M.--M.~de~Denus--Baillargeon$^{1,4}$, C.~Balkowski$^3$, A.~Boselli$^2$,
\newauthor
  K.~Fathi$^5$ and R.~C.~Kennicutt$^6$
   \\
  $^1$Laboratoire d'Astrophysique Exp\'erimentale, Observatoire du mont M\'egantic \& D\'epartement de physique, Universit\'e de Montr\'eal,\\C.P. 6128, succ. centre ville, Montr\'eal, Qu\'ebec, Canada H3C 3J7.\\
  $^2$Laboratoire d'Astrophysique de Marseille, Observatoire Astronomique Marseille--Provence, Universit\'e de Provence \& CNRS, 2 place\\Le Verrier, 13248 Marseille Cedex 4, France.\\
  $^3$Observatoire de Paris, section Meudon, GEPI, CNRS--UMR 8111 \& Universit\'e Paris 7, 5 Pl. Janssen, 92195 Meudon, France. \\
  $^4$Institut Fresnel, CNRS \& Universit\'es Aix Marseille, 13397 Marseille Cedex 20, France.\\
  $^5$Instituto de Astrofisica de Canarias C/ Via Lactea, s/n E38205 -- La Laguna (Tenerife) -- Spain.\\
  $^6$Institute of Astronomy, University of Cambridge, Madingley Road, Cambridge CB3 0HA, UK.\\
  $^7$Visiting Astronomer, Canada--France--Hawaii Telescope, operated by the National Research Council of Canada, the Centre National\\de la Recherche Scientifique de France, and the University of Hawaii.}

\date{Accepted 2007 December 14.  Received 2007 December 10; in original form
2007 October 30}
\pagerange{\pageref{firstpage}--\pageref{lastpage}} \pubyear{XXXX}

\def\LaTeX{L\kern-.36em\raise.3ex\hbox{a}\kern-.15em
    T\kern-.1667em\lower.7ex\hbox{E}\kern-.125emX}

\begin{document}

\label{firstpage}

\maketitle

\begin{abstract}
This is the second part of an \ha\ kinematics follow--up survey of
the {\it Spitzer\/} Infrared Nearby Galaxies Survey (SINGS) sample.
The aim of this program is to shed new light on the role of baryons
and their kinematics and on the dark/luminous matter relation in the
star forming regions of galaxies, in relation with studies at other
wavelengths. The data for 37 galaxies are presented. The
observations were made using Fabry--Perot interferometry with the
photon--counting camera \FM\ on 4 different telescopes, namely the
Canada--France--Hawaii 3.6m, the ESO La Silla 3.6m, the William
Herschel 4.2m, and the Observatoire du mont M\'egantic 1.6m
telescopes. The velocity fields are computed using custom IDL
routines designed for an optimal use of the data. The kinematical
parameters and rotation curves are derived using the {\it GIPSY}
software.  It is shown that non--circular motions associated with
galactic bars affect the kinematical parameters fitting and the
velocity gradient of the rotation curves. This leads to incorrect
determinations of the baryonic and dark matter distributions in the
mass models
derived from those rotation curves.  \\
\end{abstract}


\section{Introduction}

Understanding star formation mechanisms and their physical
connection to the interstellar medium (ISM) properties of galaxies
is crucial for resolving astrophysical issues such as the physical
nature of the Hubble sequence, the nature and triggering of
starbursts, and the interpretation of observations of the
high--redshift universe. These scientific objectives constitute the
core science program of the {\it Spitzer} Infrared Nearby Galaxies
Survey (SINGS) \citep{2003PASP..115..928K}.  Since the science
drivers of the project are dependent of many variables, the SINGS
project requires a comprehensive set of data including infrared
imaging and spectroscopic data, broadband imaging in the visible and
near--infrared as well as UV imaging and spectrophotometry.  The
{\it Spitzer} data and ancillary observations of the SINGS galaxies
will also provide valuable tools for understanding the physics of
galaxy formation and evolution.

The formation of individual stars from the collapse of dense
molecular clouds is relatively well known, e.g. the intensity of
star formation is strongly correlated with the column density of gas
and stars \citep{1998ApJ...498..541K}.  However, the large--scale
processes driving star formation are still poorly understood.  For
instance, the effect of gas dynamics on the regulation of star
formation is somewhat unknown.  Since young stars are often
associated with spiral arms, it is thought that protostars are
formed from compressed gas along large--scale shock fronts.
Moreover, the star formation history varies greatly along the Hubble
sequence. Elliptical galaxies, being gravitationally supported by
velocity dispersion, have exhausted their gas reservoir and hence
star--forming processes have ceased \citep{1998ARA&A..36..189K}.
Nevertheless, some elliptical galaxies having a rotating disc are
still forming stars (e.g. NGC 2974, \citealt{2007MNRAS.376.1021J}).
Spiral galaxies, on the other hand, continue to form stars and are
supported by rotation.  Understanding the effect of gas dynamics on
star formation would certainly help improving our understanding of
galaxy formation. Indeed, a Schmidt law modulated by rotation seems
to better fit the data than a simple (gas column density) Schmidt
law \citep{2003MNRAS.346.1215B, 2001ApJ...555..301M,
1989ApJ...344..685K}.

This paper presents the second part of an \ha\ kinematics survey of
the SINGS galaxies. The 37 galaxies showing \ha\ emission were
observed by means of Fabry--Perot (FP) interferometry. The paper is
organized as follows. Section \ref{observations} describes the
composition of the SINGS sample and the hardware used for the
observations.  The data reduction process, using a custom IDL
pipeline designed for an optimal use of the data, is introduced in
section \ref{reduction}. In section \ref{gipsy}, the details of the
kinematical analysis that has been done on the SINGS galaxies are
described.  Section \ref{results} presents the observational results
in the form of velocity fields, monochromatic maps,
position--velocity diagrams, and rotation curves. In section
\ref{discussion}, the effect of the bar on the observed kinematics
is discussed. Finally, the scientific applications of the
kinematical results presented in this paper are reviewed in section
\ref{conclusion}.

\section{Observations}
\label{observations}

\subsection{The sample}
\label{sample}

The sample, as described by \cite{2003PASP..115..928K}, is composed
of 75 nearby ($\Delta< 30$ Mpc, for $\mathrm{H_0}$ = 70 \kms
Mpc$^{-1}$) galaxies covering, in a three--dimensional parameter
space, a wide range of physical properties:

\begin{itemize}
 \item morphological type : associated with gas fraction, star formation rate (SFR) per unit mass, and bulge/disc structure;
 \item luminosity : associated with galaxy mass, internal velocity, and mean metallicity;
 \item FIR/optical luminosity ratio : associated with dust temperature, dust optical depth, and inclination.
\end{itemize}

In particular, there are roughly a dozen galaxies in each RC3 type
(E--S0, Sa--Sab, Sb--Sbc, Sc--Scd, Sd--Sm, Im--I0) leading to an
extensive set of combinations of luminosity and FIR/optical
luminosity ratio.  In particular, a factor of $10^5$ in infrared
luminosity and $10^3$ in $L_{FIR} / L_{opt}$ is covered by the
sample.  The 75 galaxies also represent a vast range of other galaxy
properties such as nuclear activity, surface brightness,
inclination, CO/\hi\ ratio, spiral arm structure, bar structure, and
environment. The galaxies were chosen as far as possible from the
Galactic plane in order to avoid Galactic extinction and high
density of foreground stars.

Since gas fraction is correlated with morphological type
\citep{2001AJ....121..753B}, not all the 75 galaxies are \ha\
emitters. In fact, \ha\ was not detected for ten galaxies (E--S0 and
Irr types), thus kinematical information could not be extracted for
those galaxies from emission lines.  \ha\ kinematics for 28 galaxies
of the SINGS sample have already been published in
\cite{2006MNRAS.367..469D}. This paper presents the second part of
the follow--up survey, namely the \ha\ kinematics of the remaining
37 galaxies.  Table \ref{basic_parameters} presents the basic galaxy
parameters.

\begin{table*} 
\centering
\begin{minipage}{140mm}  
\caption{\mbox{Observational data for the SINGS \ha\ kinematics sample} \label{basic_parameters} }
\begin{tabular}{ccccccccc}
\hline
\hline
Galaxy & $\alpha$(J2000) & $\delta$(J2000)  &  Type & $\Delta$ \footnote{$\Delta$: distance. Taken from Kennicutt et al. (2003).} &  $D_{25}^{b,i}$ \footnote{$D_{25}^{b,i}$: apparent major diameter at the 25 mag arcsec$^{-2}$ in B. Taken from the RC3.}   &  $B_T^{b,i}$  \footnote{$B_T^{b,i}$: total apparent magnitude in B.  Taken from the RC3.}  &  $M_B^{b,i}$  \footnote{$M_B^{b,i}$: total absolute magnitude in B. Calculated from $\Delta$ and $B_{T}^{b,i}$.} & $V_{sys}$ \footnote{$V_{sys}$: systemic velocity.  Taken from Kennicutt et al. (2003).} \\
  name &    (hh mm ss) &    (\Deg \, \Min \, \Sec) &    RC3 & (Mpc) &  (arcmin)  &    &   &   (\kms)  \\
\hline
NGC 24      & 00 09 56.7    & $-$24 57 44 & SA(s)c    &  8.2 & 5.8 &   12.19   &   $-$17.38  & 554 \\
NGC 337     & 00 59 50.3    & $-$07 34 44 & SB(s)d    & 24.7 & 2.9 &   12.06   &   $-$19.90  & 1650\\
NGC 855     & 02 14 03.6    & $+$27 52 38 & E         &  9.6 & 2.6 &   13.30   &   $-$16.61  & 610 \\
NGC 1097    & 02 46 19.0    & $-$30 16 30 & SB(r'1)b  & 16.9 & 9.3 &   10.23   &   $-$20.91  & 1275\\
NGC 1291    & 03 17 18.6    & $-$41 06 29 & SB(l)0/a  &  9.7 & 9.8 &    9.39   &   $-$21.75  & 839 \\
NGC 1482    & 03 54 39.3    & $-$20 30 09 & SA0       & 22.0 & 2.5 &   13.10   &   $-$18.61  & 1655\\
NGC 1512    & 04 03 54.3    & $-$43 20 56 & SB(r)ab   & 10.4 & 8.9 &   11.13   &   $-$18.96  & 896 \\
NGC 1566    & 04 20 00.4    & $-$54 56 16 & SAB(rs)   & 18.0 & 8.3 &   10.33   &   $-$20.95  & 1496\\
NGC 1705    & 04 54 13.5    & $-$53 21 40 & SA0       &  5.8 & 1.9 &   12.77   &   $-$16.05  & 628 \\
Ho II       & 08 19 05.0    & $+$70 43 12 & Im        &  3.5 & 7.9 &   11.10   &   $-$16.62  & 157 \\
DDO 053     & 08 34 07.2    & $+$66 10 54 & Im        &  3.5 & 1.5 &   14.70   &   $-$13.02  & 19  \\
NGC 2841    & 09 22 02.6    & $+$50 58 35 & SA(r)b    &  9.8 & 8.1 &   10.09   &   $-$19.87  & 638 \\
Ho I        & 09 40 32.3    & $+$71 10 56 & IAB(s)m   &  3.5 & 3.6 &   13.00   &   $-$14.72  & 143 \\
NGC 3034    & 09 55 52.2    & $+$69 40 47 & I0        &  3.5 &11.2 &    9.30   &   $-$18.42  & 203 \\
Ho IX       & 09 57 32.0    & $+$69 02 45 & Im        &  3.5 & 2.5 &   14.30   &   $-$13.42  & 46  \\
NGC 3190    & 10 18 05.6    & $+$21 49 55 & SA(s)a    & 17.4 & 4.4 &   12.12   &   $-$19.08  & 1271\\
IC 2574     & 10 28 21.2    & $+$68 24 43 & SAB(s)m   &  3.5 &13.2 &   10.80   &   $-$16.92  & 57  \\
NGC 3265    & 10 31 06.8    & $+$28 47 47 & E         & 20.0 & 1.3 &   13.00   &   $-$18.50  & 1421\\
Mrk 33      & 10 32 31.9    & $+$54 24 03 & Im        & 21.7 & 1.0 &   13.20   &   $-$18.50  & 1461\\
NGC 3351    & 10 43 57.7    & $+$11 42 13 & SB(r)b    &  9.3 & 7.4 &   10.53   &   $-$19.31  & 778 \\
NGC 3627    & 11 20 15.0    & $+$12 59 30 & SAB(s)b   &  8.9 & 9.1 &    9.65   &   $-$20.10  & 727 \\
NGC 3773    & 11 38 13.0    & $+$12 06 43 & SA0       & 18.3 & 1.2 &   12.90   &   $-$18.40  & 987 \\
NGC 4254    & 12 18 49.6    & $+$14 24 59 & SA(s)c    & 20.0 & 5.4 &   10.44   &   $-$21.07  & 2407\\
NGC 4450    & 12 28 29.6    & $+$17 05 06 & SA(s)ab   & 20.0 & 5.2 &   10.90   &   $-$20.61  & 1954\\
NGC 4559    & 12 35 57.7    & $+$27 57 35 & SAB(rs)cd & 11.6 &10.7 &   10.46   &   $-$19.86  & 816 \\
NGC 4594    & 12 39 59.4    & $-$11 37 23 & SA(s)a    & 13.7 & 8.7 &    8.98   &   $-$21.70  & 1091\\
NGC 4631    & 12 42 08.0    & $+$32 32 26 & SB(s)d    &  9.0 &15.5 &    9.75   &   $-$20.02  & 606 \\
NGC 4736    & 12 50 53.0    & $+$41 07 14 & SA(r)ab   &  5.3 &11.2 &    8.99   &   $-$19.63  & 308 \\
DDO 154     & 12 54 05.2    & $+$27 08 59 & IB(s)m    &  5.4 & 3.0 &   13.94   &   $-$14.72  & 376 \\
NGC 4826    & 12 56 43.7    & $+$21 40 52 & SA(rs)ab  &  5.6 &10.0 &    9.36   &   $-$19.38  & 408 \\
DDO 165     & 13 06 24.8    & $+$67 42 25 & Im        &  3.5 & 3.5 &   12.80   &   $-$14.92  & 37  \\
NGC 5033    & 13 13 27.5    & $+$36 35 38 & SA(s)c    & 13.3 &10.7 &   10.75   &   $-$19.87  & 875 \\
NGC 5408    & 14 03 20.9    & $-$41 22 40 & IB(s)m    &  4.5 & 1.6 &   12.20   &   $-$16.07  & 509 \\
NGC 5474    & 14 05 01.6    & $+$53 39 44 & SA(s)cd   &  6.9 & 4.8 &   11.28   &   $-$17.91  & 273 \\
NGC 6822    & 19 44 56.6    & $-$14 47 21 & IB(s)m    &  0.6 &15.5 &    9.31   &   $-$14.58  &$-$57\\
NGC 7552    & 23 16 11.0    & $-$42 34 59 & SB(s)ab   & 22.3 & 3.4 &   11.25   &   $-$20.49  & 1585\\
NGC 7793    & 23 57 49.8    & $-$32 35 28 & SA(s)d    &  3.2 & 9.3 &    9.63   &   $-$17.90  & 230 \\

\hline
\end{tabular}
\end{minipage}
\end{table*}

\subsection{Observing runs}
\label{hardware}

The observations have been obtained with the same instrumental
set--up consisting of a scanning Fabry--Perot (FP) interferometer,
an imaging device designed for faint fluxes, and a narrow--band
($\sim$15\AA{}) interference filter.  For imaging, a
photon--counting camera (\FM) and an Andor commercial ({\scriptsize
L3CCD}) camera were used. Each instrument was attached to a focal
reducer at the Cassegrain or Nasmyth focus of the telescope. The
focal reducers used were Panoramix at the Observatoire du mont
M\'egantic (OmM) 1.6m telescope, Cigale at the European Southern
Observatory (ESO) La Silla 3.6m telescope, MOS/FP at the
Canada--France--Hawaii 3.6m telescope (CFHT), and GHaFaS at the
William Herschel 4.2m telescope (WHT).  Table \ref{telescopes}
describes the various characteristics of the instruments.

The spectral profiles for every pixel in the field of view were
obtained by scanning the free spectral range (FSR) of the
Fabry--Perot.  The FSR is the wavelength interval between two
adjacent transmission peaks :

\begin{equation}
\label{FSR_eq}
 \mathrm{FSR=\lambda_0/p}
\end{equation}

\noindent where $\lambda_0$ is the rest wavelength and p the
interference order. The number of channels needed to scan the FSR
must be at least 2.2 times the FP \textit{finesse F} for a good
sampling (Nyquist criteria).  The \textit{finesse} is a
dimensionless parameter representing the spectral resolving power R
of the scanned line and is related to the full--width half--maximum
(FWHM) of the transmitted line:

\begin{equation}
\label{finesse_eq}
  \textit{F}= \mathrm{ \frac{R}{p} = \frac{FSR}{FWHM} }
\end{equation}

The FP etalon used for the observations has a high interference
order (typically p=765 at H$\alpha$) and is capable of achieving
high values for the \textit{finesse} and spectral resolution.  A
typical observation would be to scan the FP etalon in 48 channels
with an exposure time of 15s per channel, then repeat the process
for 15 cycles.  However, when the sky transparency is not excellent
during an exposure, the etalon is scanned more rapidly with an
exposure time of 10s per channel so that the atmospheric conditions
can be averaged out more efficiently.  The resulting data cube is a
set of interferograms stacked together, where each one represents an
image of the object modulated by the interference pattern for a
given FP spacing.

The filter set used has 24 narrow--band \ha\ filters covering the
galaxies' systemic velocities ranging from --300 to 10 0000 \kms.
Sometimes, the filter was tilted by a few degrees to adjust its
central wavelength to the Doppler shifted galaxy emission.
Narrow--band filters, used to select the proper order that will go
through the etalon, allow for the \ha\ emission to pass while at the
same time blocking most of the night sky emission.

The photon--counting cameras \FM\ I \&\ II
\citep{2003SPIE.4841.1472H} consist of a GaAs Hamamatsu
photomultiplier tube having a quantum efficiency of $\sim$23\%
coupled to a Dalsa commercial {\scriptsize CCD}.  The absence of
read--out noise for this camera enables one to scan very rapidly the
FP interferometer whereas {\scriptsize CCDs} need long exposures to
overcome their read--out noise.  Consequently, \FM\ can achieve high
signal to noise ratios (S/N) and thus is ideal for faint fluxes like
the emission found in galaxies \citep{2002PASP..114.1043G}.
Additionally, a camera using a low light level charge--coupled
device ({\scriptsize L3CCD}) made by Andor Technology was also used
as the imaging device. {\scriptsize L3CCDs} have high quantum
efficiency ($\sim$80 per cent) and sub--electron readout noise
($\sigma<$0.1 e-). This kind of sensor differs from traditional
{\scriptsize CCDs} in the sense that the signal is amplified before
it reaches the output circuitry which is the major source of noise.
Gain is created by passing electrons through a multiplication
register where an electron will create a second one by avalanche
multiplication. More details about the {\scriptsize L3CCD} can be
found in \cite{2004SPIE.5499..219D} and in
\cite{2006SPIE.6276E..42D}.

The observations of the sample were spread over nine different
observing runs over a three year period.  Five runs took place at
the OmM 1.6m telescope where \FM\ is a permanent instrument.  A
second generation instrument, called \FM\ II, has been built
recently and was tested successfully on the faint dwarf galaxy DDO
154 during one of these runs.  Also, a new instrument for the La
Silla NTT is under development and regarding this matter, the
{\scriptsize L3CCD} camera was used to test its capabilities on very
faint fluxes like galaxies.  Therefore, four galaxies were observed
with this camera during the same run at the OmM Observatory.  Two
runs took place at the ESO La Silla 3.6m telescope and one at the
CFHT 3.6m telescope, both where \FM\ I is a visitor instrument.  A
last observing run took place at the WHT 4.2m telescope with the new
instrument GHaFas \citep{2007arXiv0705.4093C}. The Fabry--Perot
observations parameters for each galaxy can be found in Table
\ref{journal}.

\begin{table}
\centering
\caption{Telescope and instrument characteristics \label{telescopes}}
\begin{tabular}{@{}cccccc@{}}
\hline
\hline
Telescope   &   Instrument& Pixel Size  & FOV    \\
\,          &           &   (arcsec)    &(arcmin) \\
\hline
OmM     &   \FM     &   1.61    &   19.43   \\
OmM     &   \FM\ II &   1.54    &   18.55   \\
OmM     &   L3CCD   &   1.07    &   12.89   \\
ESO     &   \FM     &   0.42    &   5.02        \\
CFHT    &   \FM     &   0.48    &   5.84        \\
WHT     &   GHaFaS  &   0.40    &   4.82       \\
\hline
\end{tabular}
\end{table}

\begin{table*}   
\centering
\begin{minipage}{140mm}
\caption{\mbox{Journal of the Fabry--Perot
observations.}\label{journal} }
\begin{tabular}{ccccccccccccc}
\hline
\hline

Galaxy      & Date  &$\lambda_c$ &  FWHM &$\mathrm{T_{max}}$ &  $\mathrm{t_{exp}}$  &$\mathrm{t_{ch}}$ &p&FSR&F&R&$\mathrm{n_{ch}}$&$\mathrm{step_{\lambda}}$\\
\,& &\footnote{$\lambda_c$: Non--tilted filter central wavelength at 20\Deg C (in \AA{})}&\footnote{FWHM: Non--tilted filter Full--Width Half--Maximum at 20\Deg C (in \AA{})}&\footnote{$\mathrm{T_{max}}$: Non--tilted filter transmission at $\lambda_c$ and at 20\Deg C (in \%)} &\footnote{$\mathrm{t_{exp}}$: Total exposure time (in min)}& \footnote{$\mathrm{t_{ch}}$: Total exposure time per channel (in min)}&\footnote{p: Interference order at \ha}&\footnote{FSR: Free spectral range at \ha\ (in \kms)}&\footnote{F: Finesse}&\footnote{R: Resolution according to the finesse}&\footnote{$\mathrm{n_{ch}}$: Number of FP channels}  &\footnote{$\mathrm{step_{\lambda}}$: wavelength difference between channels (in \AA)}  \\

\hline
NGC 24 \footnote{ESO: European Southern Observatory, La Silla, Chile, 3.6m telescope.\label{ESO}}   & 2005/11/04    & 6581& 19.8    &60&    150 &   2.50    &   765 &   392  & 19.7  &   15071   &   60  &   0.14    \\
NGC 337\footref{ESO}    & 2005/11/02    & 6598& 18.2&73&    150 &   2.50    &   765 &   392  &   20.5    &   15657   &   60  &   0.14    \\
NGC 855\footnote{OmM: Observatoire du mont M\'egantic, Qu\'ebec, Canada, 1.6m telescope.\label{OmM}}& 2003/11/27    &6584& 15.5 &74&    190 &   4.75    &   609 &   492  &   13.2    &   8010    &   40  &   0.27     \\
NGC 1097\footref{ESO}   & 2005/11/07    & 6598& 18.2&73&    470 &   7.83    &   765 &   392  &   20.0    &   15321   &   60  &   0.14 \\
NGC 1291\footref{ESO}   & 2005/11/06    & 6584& 15.5&74&    385 &   6.42    &   765 &   392  &   20.5    &   15669   &   60  &   0.14 \\
NGC 1482\footref{ESO}   & 2005/11/08    & 6608& 16.2&69&    115 &   1.92    &   765 &   392  &   20.3    &   15561   &   60  &   0.14 \\
NGC 1512\footref{ESO}   & 2005/11/03    & 6584& 15.5&74&    740 &   2.67    &   765 &   392  &   17.5    &   13389   &   60  &   0.14 \\
NGC 1566\footref{ESO}   & 2005/11/02    & 6598& 18.2&73&    150 &   2.50    &   765 &   392  &   20.7    &   15864   &   60  &   0.14 \\
NGC 1705\footref{ESO}   & 2005/11/03    & 6581& 19.8&60&    130 &   2.17    &   765 &   392  &   20.0    &   15282   &   60  &   0.14 \\
Ho II   \footref{OmM}   & 2005/02/05    & 6563& 30.4&80&    204 &   4.25    &   765 &   392  &   17.7    &   13527   &   48  &   0.18 \\
DDO 053\footnote{CFHT: Canada--France--Hawaii Telescope, Hawaii, USA, 3.6m telescope\label{CFHT}.}    & 2006/04/07    &6563& 30.4&80 &96  &   2.00    &   765 &   392  &   18.1    &   13885   &   48  &   0.18     \\
NGC 2841 \footref{OmM}  &2005/02/03     & 6584& 15.5&74&    240 &   5.00    &   765 &   392  &   12.7    &   9731    &   48  &   0.18     \\
Ho I    \footref{CFHT}  & 2006/04/05    & 6563& 30.4&80&    176 &   3.67    &   765 &   392  &   14.2    &   10869   &   48  &   0.18     \\
NGC 3034\footref{OmM}   & 2007/03/01    & 6581& 19.8&60&    384 &   8.00    &   765 &   392  &   17.2    &   13181   &   48  &   0.18     \\
Ho IX   \footref{OmM}   & 2005/05/10    & 6563& 30.4&80&    228 &   4.75    &   765 &   392  &   15.9    &   12185   &   48  &   0.18     \\
NGC 3190\footref{OmM}   & 2004/11/03    & 6598& 18.2&73&    144 &   3.00    &   765 &   392  &   15.6    &   11905   &   48  &   0.18     \\
IC 2574 \footref{OmM}   & 2005/02/03    & 6563& 30.4&80&    180 &   3.75    &   765 &   392  &   16.6    &   12713   &   48  &   0.18     \\
NGC 3265\footref{OmM}   & 2007/03/01    & 6598& 18.2&73&    100 &   2.08    &   765 &   392  &   18.8    &   14396   &   48  &   0.18     \\
Mrk 33  \footref{CFHT}  & 2006/04/08    & 6598& 18.2&73&    48  &   1.00    &   765 &   392  &   16.7    &   12798   &   48  &   0.18     \\
NGC 3351\footref{OmM}   & 2005/02/03    & 6584& 15.5&74&    156 &   3.25    &   765 &   392  &   18.0    &   13750   &   48  &   0.18     \\
NGC 3627\footref{OmM}   & 2005/02/06    & 6584& 15.5&74&    144 &   3.00    &   765 &   392  &   16.6    &   12670   &   48  &   0.18     \\
NGC 3773\footref{CFHT}  & 2006/04/08    & 6584& 15.5&74&    76  &   1.58    &   765 &   392  &   17.2    &   13127   &   48  &   0.18     \\
NGC 4254\footref{OmM}   & 2005/02/14    & 6621& 18.0&68&    240 &   5.00    &   765 &   392  &   17.4    &   13825   &   48  &   0.18     \\
NGC 4450\footref{ESO}   & 2002/04/07    & 6607& 12.0&69&    60  &   2.50    &   793 &   381  &   12.1    &   4604    &   24  &   0.35     \\
NGC 4559\footref{OmM}   & 2005/02/06    & 6584& 15.5&74&    408 &   8.50    &   765 &   391  &   16.5    &   12631   &   48  &   0.18     \\
NGC 4594\footnote{WHT: William Herschel Telescope, La Palma, Spain, 4.2m telescope.\label{WHT}} &  2007/07/05 & 6585 & 15.5 & 75 & 64  &  5.00 &   765&392  & 16.5  & 12623 &48 & 0.18\\
NGC 4631\footref{OmM}   & 2005/02/01    & 6584& 15.5&74&    180 &   3.75    &   765 &   392  &   18.0    &   13757   &   48  &   0.18     \\
NGC 4736\footref{OmM}   & 2005/05/11    & 6563& 30.4&80&    216 &   4.50    &   765 &   392  &   16.7    &   12745   &   48  &   0.18     \\
DDO 154 \footref{OmM}   & 2007/02/22    & 6581& 19.8&60&    320 &   8.00    &   765 &   392  &   17.1    &   13097   &   40  &   0.21     \\
NGC 4826\footref{CFHT}  & 2006/04/07    & 6563& 30.4&80&    128 &   2.67    &   765 &   392  &   17.2    &   13121   &   48  &   0.18     \\
DDO 165\footref{CFHT}   & 2006/04/06    & 6563& 30.4&80&    172 &   3.58    &   765 &   392  &   17.1    &   13091   &   48  &   0.18     \\
NGC 5033\footref{OmM}   & 2005/05/10    & 6584& 15.5&74&    460 &   9.58    &   765 &   392  &   16.7    &   12782   &   48  &   0.18     \\
NGC 5408\footref{CFHT}  & 2006/04/07    & 6581& 19.8&60&    108 &   2.25    &   765 &   392  &   17.1    &   13083   &   48  &   0.18     \\
NGC 5474\footref{OmM}   & 2007/02/28    & 6581& 19.8&60&    108 &   2.25    &   765 &   392  &   17.2    &   13157   &   48  &   0.18     \\
NGC 6822\footref{ESO}   & 2005/11/08    & 6563& 30.4&80&    60  &   1.00    &   765 &   392  &   19.8    &   15121   &   60  &   0.14     \\
NGC 7552\footref{ESO}   & 2005/11/02    & 6598& 18.2&73&    120 &   2.00    &   765 &   392  &   19.8    &   15160   &   60  &   0.14     \\
NGC 7793\footref{ESO}   & 2005/11/08    & 6563& 30.4&80&    100 &   1.67    &   765 &   392  &   19.7    &   15040   &   60  &   0.14     \\
\hline
\end{tabular}
\end{minipage}
\end{table*}

\section{Data reduction}
\label{reduction}

This section introduces the few steps towards obtaining radial
velocities and monochromatic maps from raw interferograms.  In
particular,

\begin{itemize}
 \item wavelength calibration;
 \item spectral smoothing and sky emission subtraction;
 \item adaptive spatial binning and map extraction;
 \item {\scriptsize WCS} astrometry.
\end{itemize}

For a more complete description of the data reduction steps, we
refer to \cite{2005MNRAS.360.1201H}, \cite{2006MNRAS.367..469D},
\cite{2006MNRAS.368.1016D}, and \cite{2006MNRAS.366..812C}. The
software used can be found at
http://www.astro.umontreal.ca/fantomm/reduction.

\subsection{Wavelength calibration}
\label{calib}

The raw data cube, obtained during an acquisition, must be
wavelength corrected since the transmitted wavelength
$\mathrm{\lambda}$ is a function of the angle $\mathrm{\theta}$ of
the incoming light beam :

\begin{equation}
\label{FP_eq}
  \mathrm{p \lambda = 2ne\, cos\theta }
\end{equation}

\noindent where p is the interference order at
$\mathrm{\lambda_0}$(6562.78 \AA{}), n the index of the medium, and
e the distance between the parallel plates of the etalon.  The
wavelength calibration is made by scanning the neon line at 6598.95
\AA{} just before and after a three hour acquisition, in the same
conditions as the observation itself.  This enables one to calculate
the phase shift needed to assign a wavelength to a particular FP
spacing, for every pixel of the field. This phase map transforms raw
interferograms into a wavelength--sorted data cube.  Since one can
only know the transmitted wavelength value $\pm$ FSR, an uncertainty
remains on the zero--point of the velocity scale. Comparison with
other kinematical work will remove this uncertainty. Note that the
data cubes are not flux--calibrated.

\subsection{Spectral smoothing and sky emission subtraction}
\label{sky}

A Hanning smoothing was performed on every spectrum of the
wavelength--sorted data cubes in order to remove any artifacts
caused by the discrete sampling.  After that, the strong night sky
emission lines were subtracted.  The method used for this is to
reconstruct a sky cube using the sky dominated regions and
interpolating it in the galaxy region.  This sky cube was then
subtracted from the data cube. This method has proven to be very
successful at eliminating sky residuals compared with subtracting a
median sky spectrum where both spatial and spectral inhomogeneities
in the interference filter can lead to high sky residuals.

\subsection{Adaptive spatial binning and map extraction}
\label{binning} In order to increase the signal--to--noise ratio
({\scriptsize S/N}) in the diffuse and/or faint interstellar regions
(e.g. inter--arm regions) of the observed galaxies, an adaptive
spatial binning was applied to the data cubes. This technique is
based on Voronoi diagrams where pixels are accreted into bins until
the desired {\scriptsize S/N} ($\sim$5, typically chosen) is
reached. In high intensity emission regions like the galactic center
and spiral arms, the {\scriptsize S/N} exceeds by far the targeted
minimal value and hence the high spatial resolution is maintained.
This is an improvement over the usual gaussian smoothing where the
kernel convolution would dilute the signal in high {\scriptsize S/N}
regions.

Next, the final steps are the integration of the flux under the \ha\
line for every bin, yielding a monochromatic map, and the barycenter
computation, yielding a velocity field, following the procedure
described by \cite{2006MNRAS.368.1016D}.  The radial velocities (RV)
are given in the heliocentric rest frame. A continuum map and a
velocity dispersion map are also computed by the algorithm. The
determination of the continuum threshold is a critical step in the
data reduction as it defines the position of the barycenter of the
line.  This is done by iterative procedures.

\subsection{WCS  astrometry}
\label{astrometry}

Finally, {\scriptsize WCS} coordinates are attached to the computed
maps using the task koords in the {\scriptsize KARMA} package
\citep{1996ASPC..101...80G}. No references to the World Coordinate
System are obtained during the acquisition and are necessary since
the major axes position angle (PA) of the galaxy is
field--orientation dependent. Besides, {\scriptsize WCS} astrometry
is needed to combine the \ha\ kinematics with ancillary {\scriptsize
SINGS} surveys.  The coordinates are added by comparing positions of
stars between a reference file (a redband DSS image for instance)
and the FP non--binned continuum image.

\section{kinematical parameters fitting}
\label{gipsy}

Rotation curves are computed using the task \textit{rotcur}
available in the \textit{GIPSY} software
\citep{2001ASPC..238..358V}. \textit{Rotcur} derives the kinematical
parameters for a particular galaxy by fitting tilted--rings to the
observed velocity field. More precisely, a least--squares--fitting
is done to the function :

\begin{equation}
    \label{vrot_eq}
    V_{obs}(x,y) = V_{sys} + V_{rot }(r) \, cos \, \theta \, sin \, i+V_{exp}(r) \, sin \, \theta \,sin \, i
\end{equation}

Here $V_{obs}(x,y)$ denotes the radial velocity at the pixel
coordinates (x,y), $V_{sys}$ the systemic velocity, $V_{rot }(r)$
the rotational velocity for  the corresponding radius r, $\theta$
the azimuthal angle from the major axis in the plane of the galaxy,
$i$ the inclination angle of the galaxy, and $V_{exp}$ the expansion
velocity.  Since \textit{GIPSY} does not take into account the field
rotation of the supplied velocity field with respect to {\scriptsize
WCS} coordinates, the velocity fields were all rotated in order to
compute accurate position angles (PAs) of the kinematical major
axis.

Fitting the kinematical parameters was done in a three--step
process. First, the systemic velocity $V_{sys}$ and the kinematical
center ($x_{pos},y_{pos}$) are fitted while keeping fixed the
inclination i and position angle PA.  The starting parameters used
are the photometric i and PA given by the {\it hyperleda} catalog
and for $V_{sys}$, the value used for selecting the interference
filter. The starting values for the galactic center are the
photometric center corresponding to the maximum value near the
galactic center in the continuum map for spiral galaxies or in the
\textit{Spitzer} 3.6 $\umu$m image for distorted and irregular
galaxies.  Then, a second fitting is done by letting i, PA, and
$V_{rot }$ vary with radius while keeping fixed the new--found
values for $V_{sys}$, $x_{pos}$, and, $y_{pos}$.  Finally,
\textit{rotcur} is run again, with only $V_{rot}$ as the varying
parameter.  The computed rotation curve is thus derived with fixed
kinematical parameters so that the sample could be homogeneous, even
though some velocity fields (e.g. NGC 3627) were better modeled with
parameters varying with radius.

The ring width used in the fitting procedure was set to be greater
than 3 times the pixel width for a good sampling. The result is
5\arcsec\ for the OmM observations and 2\arcsec\ for the ESO, CFHT
and WHT observations. Also, additional \textit{rotcur} runs were
done for the approaching and receding sides separately in order to
model any asymmetries arising between the two sides. Finally, the
expansion velocity $V_{exp}$ was fixed to zero for all galaxies,
thus assuming pure circular rotation.  Letting $V_{exp}$ vary with
radius did not change significantly the rotation velocities in the
computed regions.

Two \textit{rotcur} parameters remain to be explained, namely the
free angle and the weighting function. To diminish the importance of
deprojection errors, radial velocities in an opening angle (called
the free angle) of typically 35\Deg\ about the minor axis were
rejected from the least--square fitting and a $cos(\theta)$ (where
$\theta$ = angle from the major axis) weighting function was applied
to give more importance to points near the major axis. Naturally,
for more face--on galaxies, the projected velocities along the
line--of--sight possess little information about rotational
velocities resulting in large errors for the kinematical parameters.

Afterwards, the \textit{GIPSY} output was analyzed by IDL routines
(mainly computing averages and standard deviations).  The
determination of constant values for i and PA throughout the galaxy
is performed by eliminating radii where non--circular motions and
warps can occur.  For a non--barred galaxy, this results in
discarding radii located in the galactic center and outer spiral
arms where one usually finds a substantial scatter in i and PA
values. For a barred galaxy, this is complicated by the fact that
important non--circular motions are present in the bar region,
therefore this region was excluded from the fit.  Also, for barred
and non--barred galaxies, rings containing too few points or
dominated by one side of the galaxy were discarded.

\section{Results}
\label{results} This section presents the \ha\ monochromatic and RV
maps as well as the rotation curves for each galaxy in order of
increasing right ascension.  Appendix A briefly describes the
observed kinematics for each galaxy of the sample.  In Appendix B,
four images per galaxy are first shown : DSS blue image when
available (top--left), \textit{Spitzer} 3.6 $\umu$m image
(top--right), \ha\ monochromatic image (middle--left), and the
corresponding \ha\ velocity field (middle--right).  The blue images
show the intermediate ($\sim 10^9$ yrs) stellar population and the
3.6 $\umu$m the old stellar population, tracer of the total mass of
galaxies, while the \ha\ monochromatic maps show the gas ionized by
massive (M $> 8$ \msol), young ($\sim 10^6$ yrs) OB stars
\citep{1998ARA&A..36..189K, 2001AJ....121..753B}. These four maps
are WCS--oriented and represent the same field of view. For each
galaxy, the field of view is adjusted so that the \ha\ morphology
and kinematics as well as the large scale stellar morphology are
displayed with great detail. The maps and the rotation curves can be
found at http://www.astro.umontreal.ca/fantomm/singsII/

Moreover, a position--velocity (PV) diagram is provided for each
galaxy when the extraction of the kinematical parameters was
possible. They represent a slice in the data cube along the
kinematical major axis. The black line, superposed on the diagrams,
represents a cut in the model velocity field. A data cube slice
along the kinematical major axis showing a rotation curve which
superposes well on the velocities of the maximum intensity of the
ionized gas emission implies that there are no significant
non--circular motions nor any kinematical twist. Lastly, rotation
curves can be found in Appendix C.  The definition of the errors
used is either the largest velocity difference between both sides
and the approaching or receding side separately, or the
\textit{rotcur} intrinsic error if it is greater.

\begin{figure}
\centering
\includegraphics{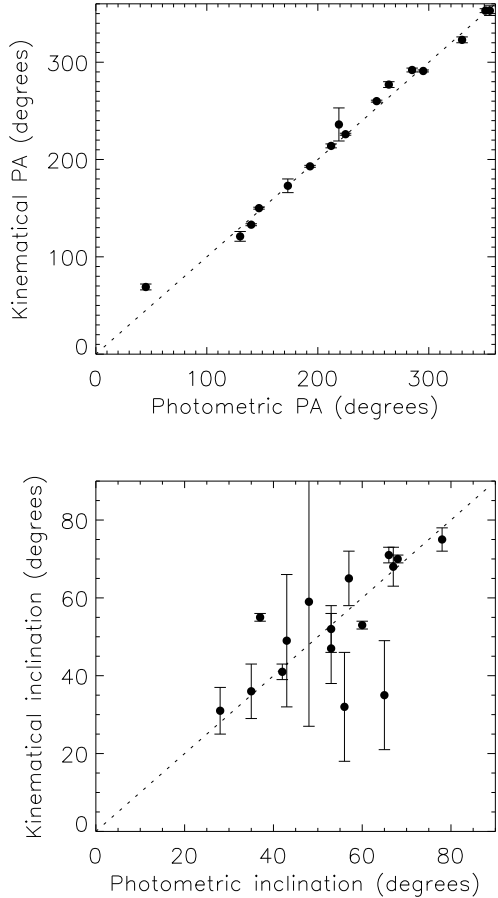}
\caption{Comparison between photometric and kinematical parameters.
Top: Position Angle.  Bottom: Inclination.  The dotted line
represents agreeing parameters.  \label{fig:inclpa}}
\end{figure}

The kinematical parameters inclination and position angle found by
the technique described in section \ref{gipsy} are presented in
Table \ref{kinparam}.  Figure \ref{fig:inclpa} compares the
kinematical parameters with respect to the photometric values. As
expected, the agreement is better for the PAs than for the
inclinations. The two points that stand out are NGC 1512 (i$_{phot}$
= 65\Deg\ \& i$_{kin}$ = 35\Deg) and NGC 1566 (i$_{phot}$ = 56\Deg\
\& i$_{kin}$ = 32\Deg). In both cases, the explanation is quite
clear: the photometric parameters are mainly representative of the
bar which contributes a large part of the light resulting in more
edge--on values. Only looking at the B or 3.6-$\umu$m images in
Figures B7 \& B8, one can see that the outer isophotes are much more
face--on.

Finally, on the overall 37 galaxies exhibiting \ha\ emission, it was
not possible to extract rotation curves for 21 of them due to either
poor spatial coverage, absence of large--scale rotation or extremely
perturbed discs.

\begin{table}
\centering
\begin{minipage}{0.5\textwidth}
\centering
\caption{Photometric and kinematical parameters \label{kinparam}}
\begin{tabular}{@{}ccccc@{}}
\hline
\hline
Galaxy      & \multicolumn{2}{|c|}{Phot} & \multicolumn{2}{|c|}{Kin}  \\
name        & PA            & i &   PA      & i     \\
\hline
NGC 24      & 225       & 78        & 226 $\pm$ 1       & 75 $\pm$ 3 \\
NGC 337     & 130       & 53        & 121 $\pm$ 5       & 52 $\pm$ 6  \\
NGC 1097    & 140       & 37        & 133 $\pm$ 1       & 55 $\pm$ 1  \\
NGC 1512    & 253       & 65        & 260 $\pm$ 1       & 35 $\pm$ 14   \\
NGC 1566    & 212       & 56        & 214 $\pm$ 2       & 32 $\pm$ 14 \\
NGC 2841    & 147       & 68        & 150 $\pm$ 1       & 70 $\pm$ 1 \\
NGC 3351    & 193       & 42        & 193 $\pm$ 1       & 41 $\pm$ 2 \\
NGC 3627    & 173       & 57        & 173 $\pm$ 7       & 65 $\pm$ 7   \\
NGC 4254    & 45        & 28        & 69  $\pm$ 3       & 31 $\pm$ 6    \\
NGC 4450    & 355       & 43        & 353 $\pm$ 5       & 49 $\pm$ 17 \\
NGC 4559    & 330       & 67        & 323 $\pm$ 3       & 68 $\pm$ 5 \\
NGC 4736    & 285       & 35        & 292 $\pm$ 2       & 36 $\pm$ 7 \\
DDO 154     & 219       & 48        & 236 $\pm$ 17      & 59 $\pm$ 32\\
NGC 4826    & 295       & 60        & 291 $\pm$ 1       & 53 $\pm$ 1 \\
NGC 5033    & 351       & 66        & 353 $\pm$ 2       & 71 $\pm$ 2 \\
NGC 7793    & 264       & 53        & 277 $\pm$ 3       & 47 $\pm$ 9 \\
\hline
\end{tabular}
\end{minipage}
\end{table}

\section{Discussion}
\label{discussion}

In this paper, the rotation curves given in Appendix \ref{app:rc}
were obtained from the kinematical parameters derived using
tilted--ring models which assume pure circular motions.  Even if we
tried to avoid the zones obviously affected by non--circular
motions, there is still considerable work needed to extract rotation
curves that are truly representative of the mass distribution, and
can be used for mass modeling purposes.  This is especially true for
barred systems, which account for about one third of the galaxies in
Table \ref{basic_parameters}.

The idea behind obtaining an accurate determination of the
gravitational potential is correlated with the dark halo modeling.
For instance, there has been significant debate about the shape of
dark matter density profiles, especially regarding their inner
slope. Based on cosmological N--body simulations (Navarro et al.
1996, 1997, hereafter collectively NFW), the dark matter halo
profile appears to be independent of halo mass with an inner
logarithmic slope equal to $-1$. Nevertheless, recent higher
resolution simulations suggest that the density profiles do not
converge to a single power law at small radii. At the smallest
resolved scales (0.5\% of the virial radius), profiles usually have
slopes between --1 and --1.5 (Moore et al. 1999; Ghigna et al. 2000;
Jing \& Suto 2000; Fukushige \& Makino 2001; Klypin et al. 2001;
Power et al. 2003; Navarro et al. 2004; Diemand et al. 2004).

In addition, all simulations find density profiles that are
inconsistent with the isothermal profile found in observations.  In
the outer regions, the determination of the dark halo slope based on
mapping the outer density profile of galaxies is difficult, owing
mainly to a lack of mass tracers at large radii.  In the inner
regions, the unknown value of the stellar mass--to--light ratio
further complicates the determination of the mass distribution. This
has led to dedicated analysis on dwarf and low surface brightness
(LSB) galaxies that are believed to be dark matter dominated at all
radii (de Blok \& McGaugh 1997; Verheijen 1997; Swaters 1999).  It
has been suggested that rotation curves of dwarf and LSB galaxies
rise less steeply than predicted by numerical simulations based on
the cold dark matter (CDM) paradigm (Moore 1994; Flores \& Primack
1994; de Blok \& McGaugh 1997; McGaugh \& de Blok 1998; de Blok et
al. 2001a, 2001b).

However, a number of observational uncertainties cast doubt over
these early claims.  These include beam smearing for \hi\ rotation
curves (Swaters et al. 2000; van den Bosch et al. 2000), high
inclination angles and \ha\ long--slit alignment errors (Swaters et
al. 2003a), and non--circular motions close to the center of
galaxies (Swaters et al. 2003b). Many of these uncertainties can be
quantified or eliminated by measuring high--resolution
two--dimensional velocity fields (Barnes et al. 2004). At optical
wavelengths, these can be obtained via Fabry--Perot interferometry
(e.g., Blais--Ouellette et al. 1999) or integral field spectroscopy
(e.g., Andersen \& Bershady 2003; Courteau et al. 2003).

There are ways to extract the true kinematics that reflect the
gravitational potential. One is to derive the potential directly
from the 2D velocity field or the 3D data cube (this is work in
preparation). The other way is to derive the bar parameters using
the {\it Spitzer} images, compare with numerical simulations, and
apply the necessary corrections for, e.g. the streaming motions
induced by the bars (see Hernandez et al. 2005b; Perez, Fux \&
Freeman, 2004). This detailed work will be done in another paper
(Hernandez et al. 2007 in preparation), but we can illustrate what
needs to be done by using three of the barred systems in our sample.

\begin{figure}
\centering
\includegraphics[width=0.45\textwidth]{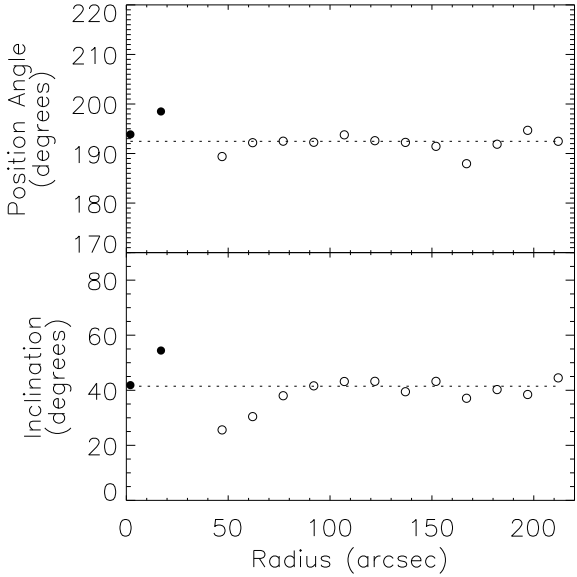}
\caption{Tilted--ring model for NGC 3351.  The dotted line shows the
fitted PA (193\Deg) and inclination (41\Deg).  The stellar bar ends
at a radius of 47\arcsec. So, the filled circles give the values
inside the bar and the empty circles the values outside the bar
region.\label{fig:modelM95}}
\end{figure}

The galaxy NGC 3351, better known as Messier 95, is a SBb galaxy,
member of the Leo group.  Situated at a distance of 9.3 Mpc, this
starburst galaxy has a large--scale stellar bar which has a
deprojected value of 47\arcsec\ for the semi--major axis
\citep{1995AJ....109.2428M}.  Outside this bar, the \ha\ velocity
field is fairly regular and the kinematical PA and inclination found
by {\textit GIPSY} agree well with the photometric values.  Thus,
the gas outside the bar is thought to be in circular orbits so that
the rotation curve represents accurately the kinematics of this
galaxy. The small deviations from circular motions are due to
streaming along the inner ring.  The \ha\ rotation curve shows a
more or less constant velocity beginning at the end of the stellar
bar with peaks corresponding to the 70\arcsec\ inner ring.  Figure
\ref{fig:modelM95} displays the tilted--ring model results and one
can see, outside the bar, the fairly regular values for the position
angle and inclination as a function of the radius.

However, inside the large--scale bar, the \ha\ velocity field shows
the perturbed kinematics expected for this barred system. The
twisted isovelocity contours indicate that a pure circular rotation
model will present a poor fit to the data and hence kinematical
information extracted in this region will be incorrect. This is
illustrated in Figure \ref{fig:modelM95} where the filled circles
represent the values computed in the bar region.  It is thus
imperative to take into account the bar location when deriving
kinematical parameters.

Furthermore, bar modeling is crucial in order to properly compute
the rotation curve in this region.  Numerical computations were
performed with \texttt{GADGET}, a tree--based N--body$+$\textsc{SPH}
code developed by Springel, Yoshida \& White (2001).  For the needs
of the simulations, an initial stellar population is set up to
reproduce a disc galaxy with an already formed bulge.  The initial
positions and velocities of the stellar particles are drawn from a
superposition of two axisymmetric Miyamoto--Nagai discs (Miyamoto \&
Nagai 1975) of mass respectively $10^{10}$ and $10^{11}$\msol, of
scale lengths respectively $1$ and $3.5$~kpc and common scale height
of $0.5$~kpc. Velocity dispersions are computed solving numerically
the Jeans equations.  The total number of stellar particles is
1.1~$\times 10^6$. The run includes a dark halo made of 2.2~$\times
10^6$ live particles distributed in a Plummer sphere of scalelength
$50$~kpc and of mass respectively 2.42 and 6.46~$\times
10^{11}$\msol. The total mass of the gas is 0.11~$\times
10^{11}$\msol. Finally, the total mass of the simulated galaxy is
7.67~$\times 10^{11}$\msol.

\begin{figure*}
\begin{center}
\includegraphics[width=5.7cm]{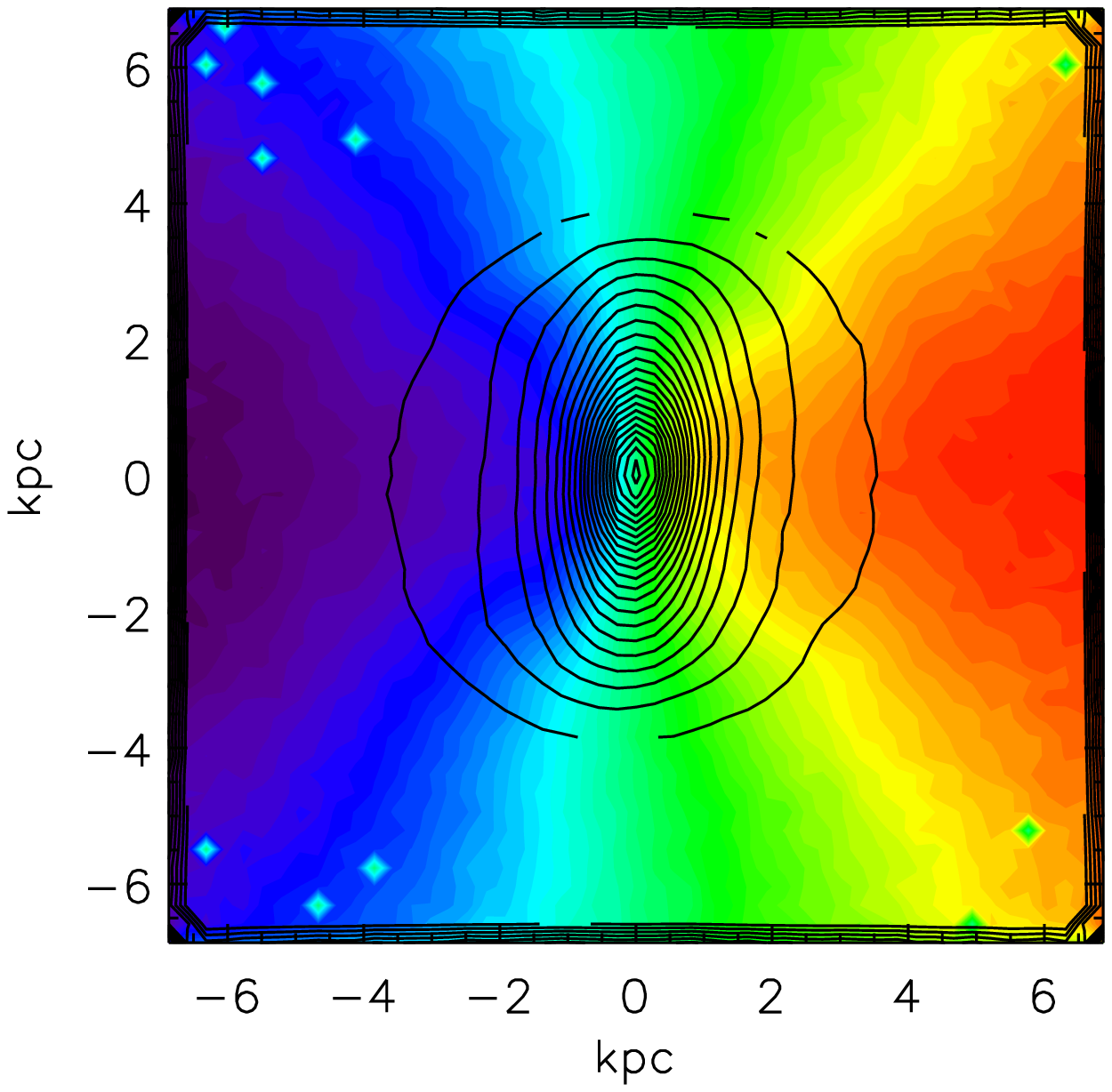}
\includegraphics[width=5.2cm]{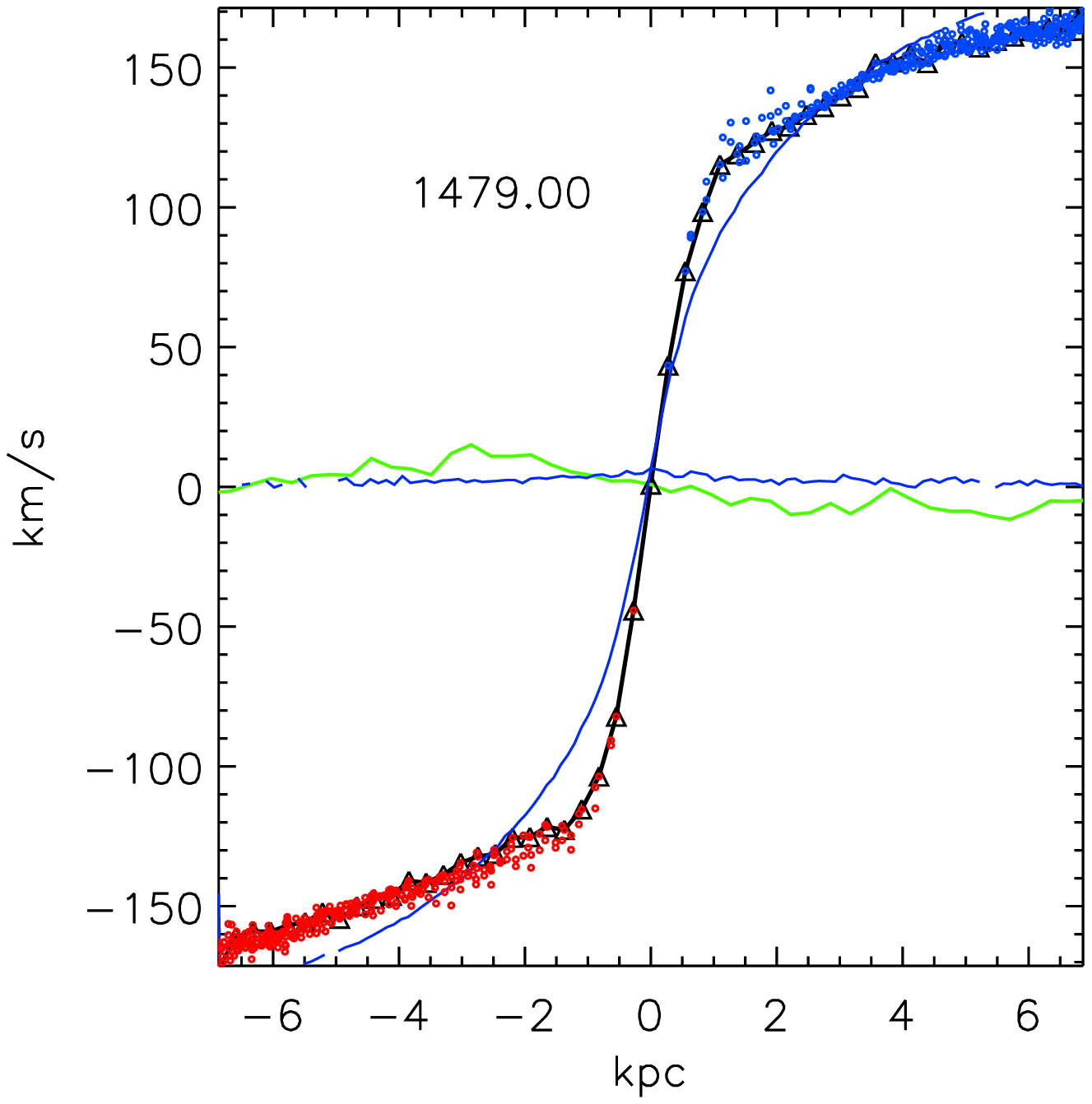}
\includegraphics[width=7.4cm]{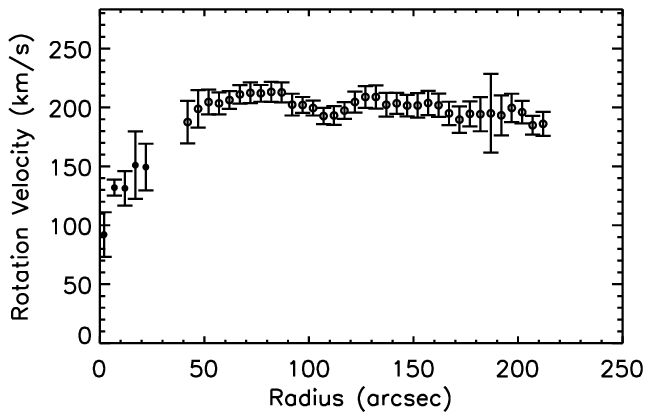}\\
\includegraphics[width=5.7cm]{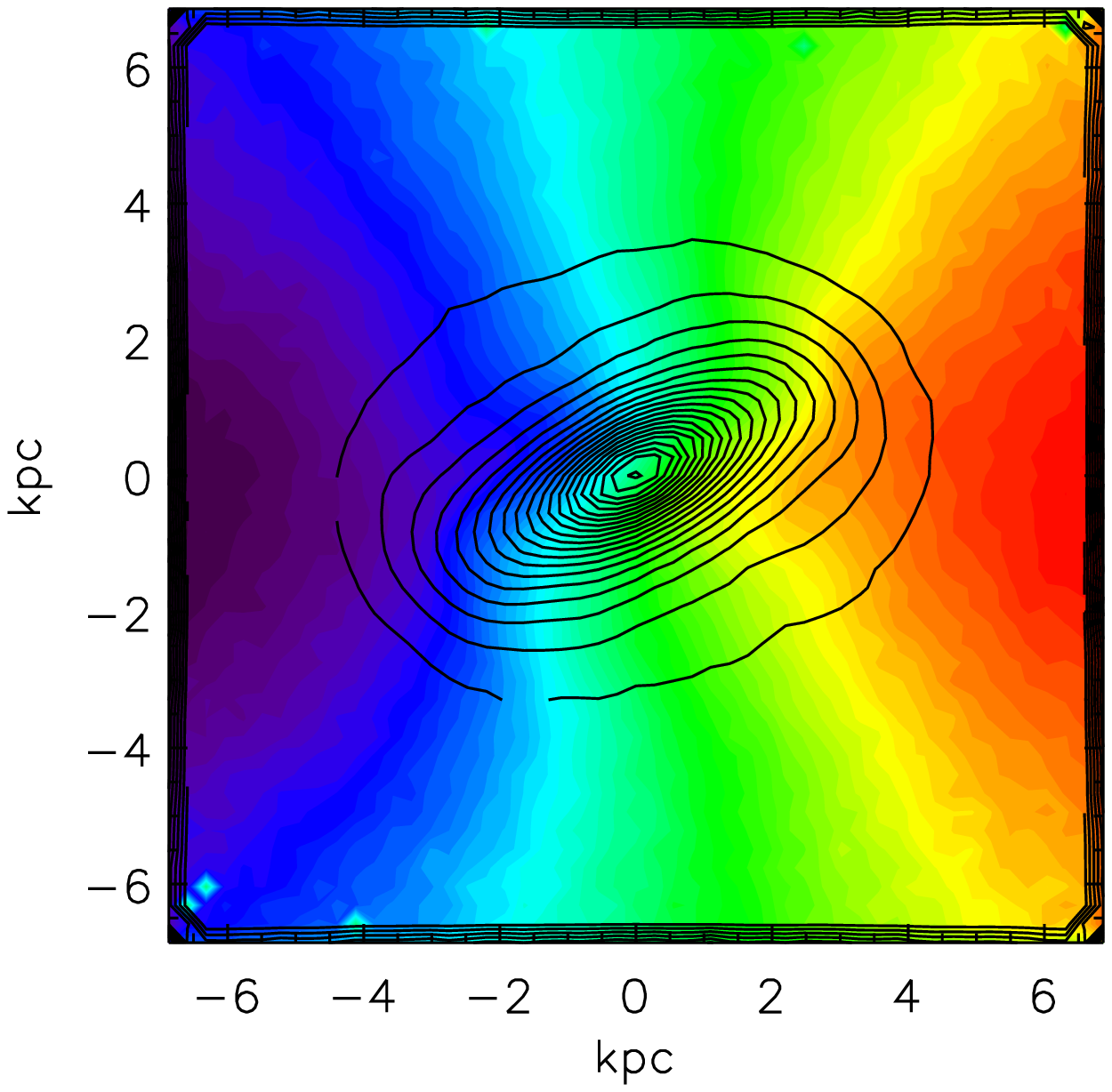}
\includegraphics[width=5.2cm]{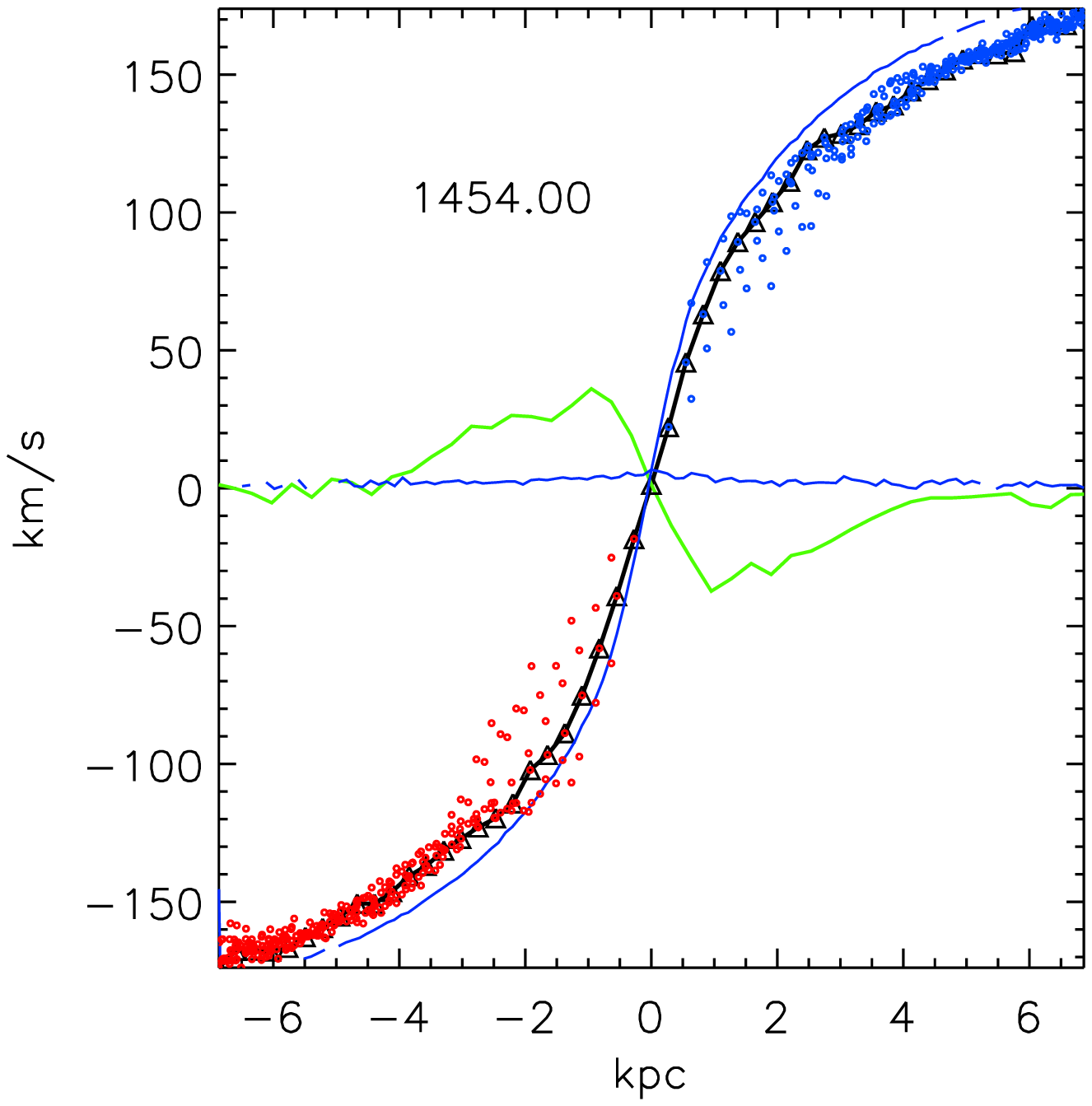}
\includegraphics[width=7.4cm]{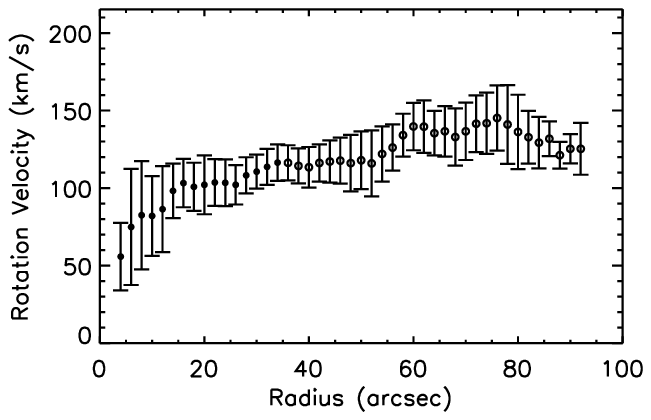}\\
\includegraphics[width=5.7cm]{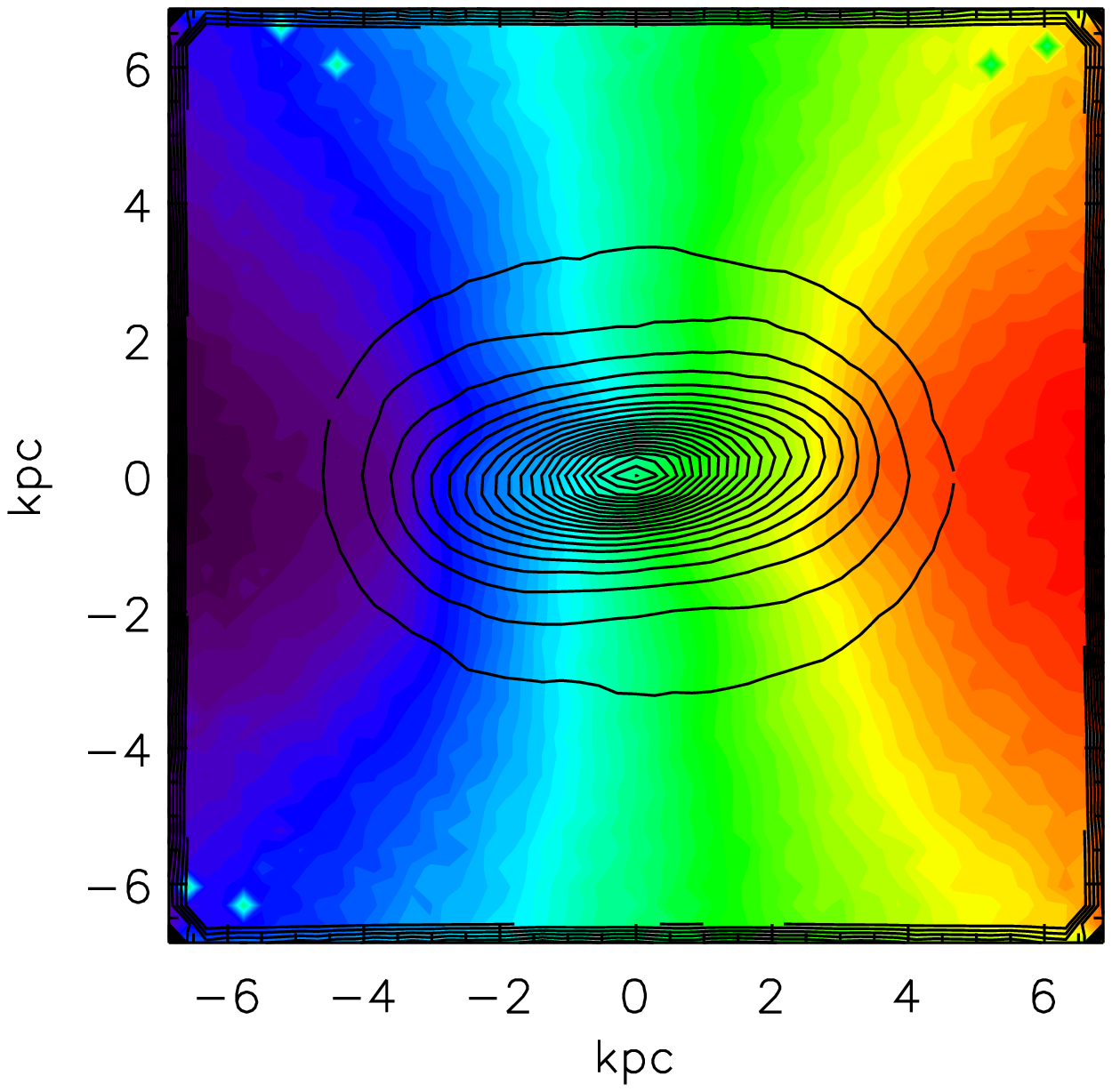}
\includegraphics[width=5.2cm]{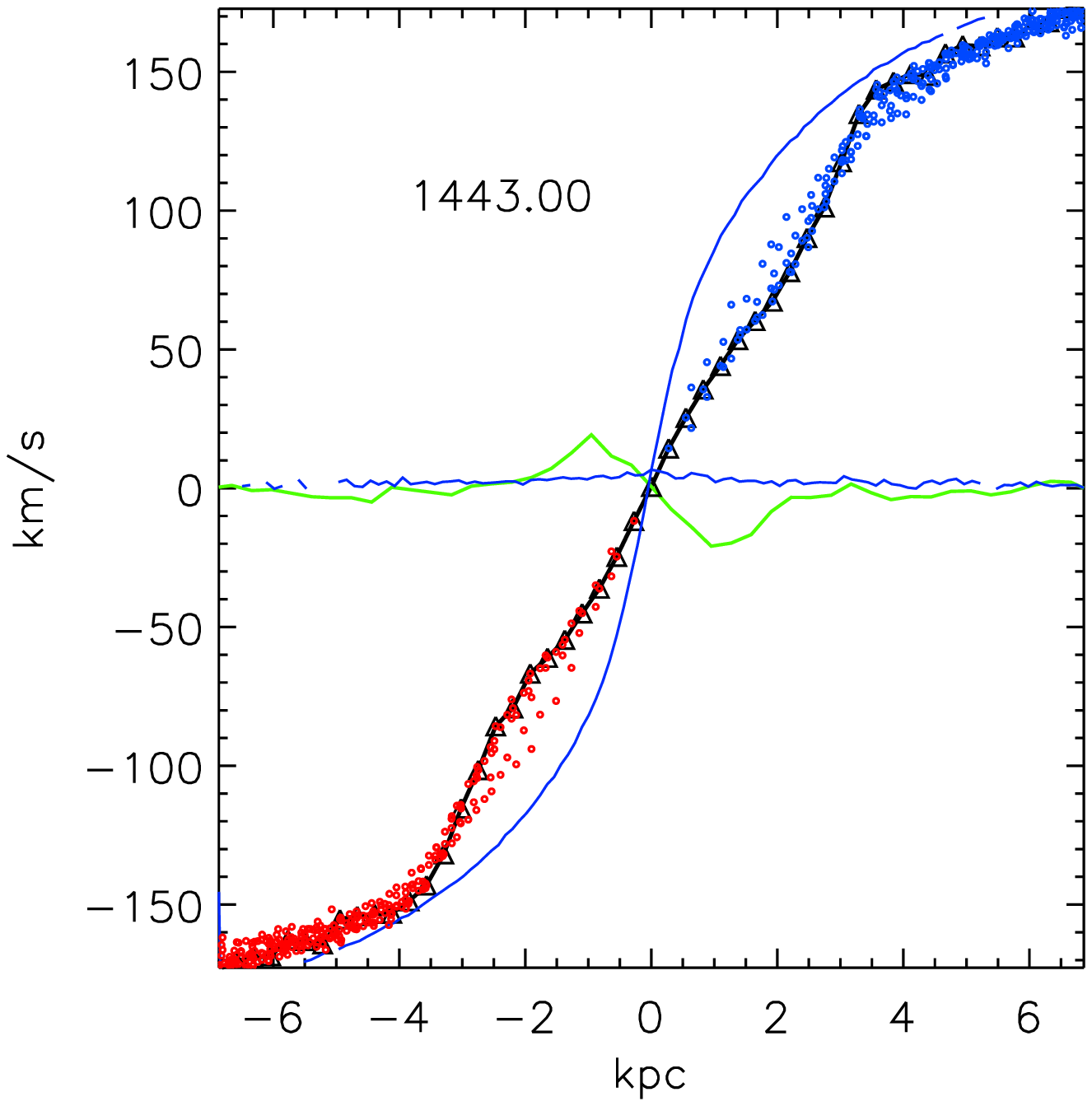}
\includegraphics[width=7.4cm]{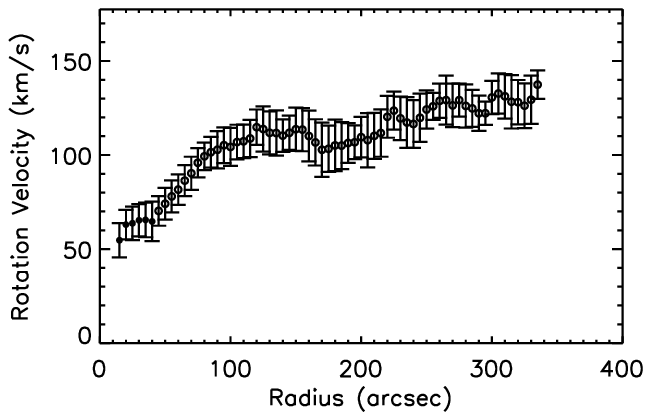}
\caption{Three different bar orientations with respect to the major
axis. Top: NGC 3351, perpendicular position.  Middle: NGC 337,
intermediate position.  Bottom: NGC 4559, parallel position. (left)
Density contours of the bar superposed on the velocity field of the
model. (middle) The thin blue and green lines represent the rotation
velocities along the major and minor axis, respectively, for an
input model with pure rotation. The dots (blue for the approaching
side and red for the receding side) represent the observed
velocities. (right) Rotation curves derived in this study.
\label{fig:OH}}
\end{center}
\end{figure*}

Numerical simulations of the kinematical effect of the bar are shown
in Figure \ref{fig:OH}.  The input model for the galaxy is pure
rotation and the corresponding rotation velocities along the major
and minor axis are shown in faint blue and green lines.  Afterwards,
the code simulates galaxy evolution where a bar is developing.  The
difference in orientation between the major axis and the bar
corresponds to different evolving times which are given in units of
millions years.  For NGC 3351, the kinematical PA
($PA_{kin}=193$\Deg) is almost perpendicular to the bar, with
$PA_{bar} = 113$\Deg\ using the value found by
\cite{2007ApJ...657..790M}.  The results of the simulations are
shown in the form of RV maps and stellar densities superposed.  For
an 80\Deg\ difference between the major axis and the bar PAs, the
overall effect is an artificial increase in the velocity gradient.
This can be explained by gas moving along x1 orbits (parallel to the
bar) where the velocity is greater at the perigalacticon (near the
center) than at the apogalacticon (near the end of the bar).  The
true kinematics of this galaxy will therefore be obtained by
correcting this artificial increase in the velocity gradient.  Until
then, rotation velocities inside the bar region are not given for
the final rotation curve.  See the galaxy description in Appendix A
for additional evidences of non--circular motions.

\begin{figure}
\begin{center}
\includegraphics[width=0.45\textwidth]{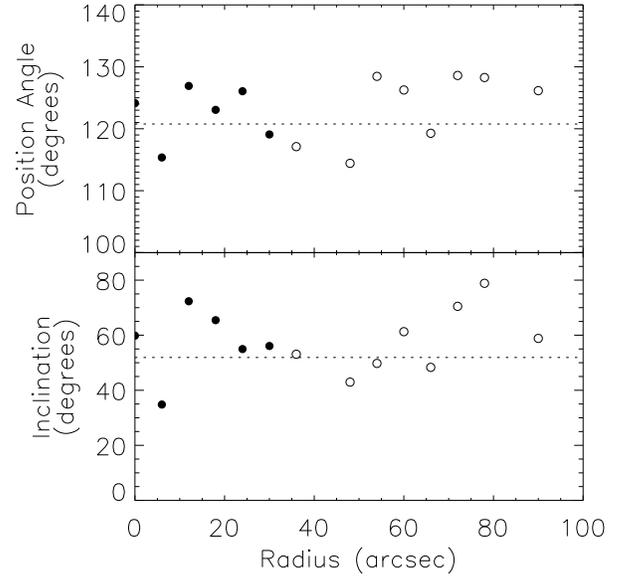}
\caption{Tilted--ring model for NGC 337.  The dotted line shows the
fitted PA (121\Deg) and inclination (52\Deg). The stellar bar ends
at a radius of 35\arcsec. So, the filled circles give the values
inside the bar and the empty circles the values outside the bar
region.\label{fig:modeln337}}
\end{center}
\end{figure}

Another galaxy displaying a perturbed velocity field is NGC 337.
This asymmetric SBd galaxy features an off--center stellar bar
having a PA of 162\Deg\ and a deprojected value of 35\arcsec\ for
the bar semi--major axis \citep{2007ApJ...657..790M}.  Since the
photometric major axis of this galaxy is 130\Deg, the bar has an
intermediate orientation with respect to the major axis.  Numerical
simulations have been done for this bar position and the results are
illustrated in the middle panel of Figure \ref{fig:OH}.  The
simulated velocity field displays the characteristic Z--shape of the
isocontours similar to what is seen in the \ha\ velocity map.  One
characteristic is the velocity gradient along the minor axis (green
line in Figure \ref{fig:OH}). Another feature is the relative
agreement between the input rotation velocities (faint blue line)
and the computed ones (triangle symbols show averaged red and blue
points calculated using the method presented in section
\ref{gipsy}). The rotation curve in the bar region is thus provided
for this galaxy since these perturbations are confined to the minor
axis which is excluded from the fit.  The tilted--ring model for
this galaxy, presented in Figure \ref{fig:modeln337}, illustrates
the small differences for the fitted values between the bar region
(filled circles) and spiral arms (open circles).

After having completed the kinematical fitting procedure, one can
look at the {\it Spitzer} IRAC 3.6$\umu$m image in order to compare
the bar location with the velocity residuals. Since the eastern side
is approaching and the western side is receding, the spiral arms
seem to be trailing in an anti--clockwise direction.  The velocity
residuals from the pure circular rotation model indicate positive
residuals east of the bar as well as negative residuals on the
southern end of the bar, suggesting an inflow of gas towards the
bar.

Two other barred galaxies having a bar orientation intermediate to
the major axis are NGC 1566 and NGC 3627, but their Seyfert activity
and warped disc respectively make difficult the kinematical analysis
regarding the effects of the bar.  Their rotation curves are
provided at the end of this paper.

\begin{figure}
\centering
\includegraphics[width=0.45\textwidth]{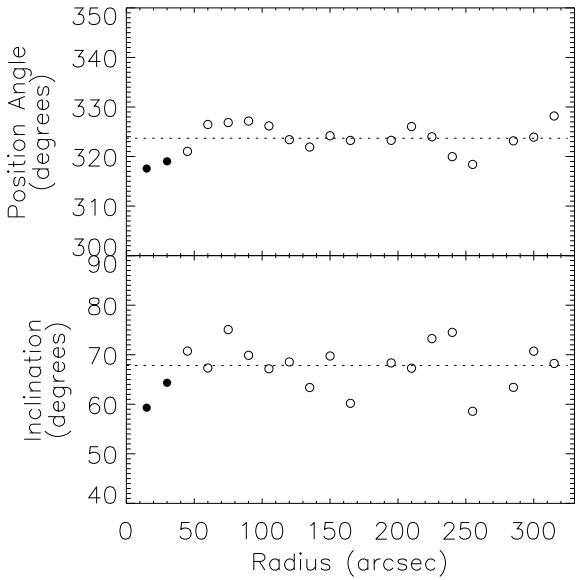}
\caption{Tilted--ring model for NGC 4559.  The dotted line shows the
fitted PA (323\Deg) and inclination (68\Deg) values.  The stellar
bar ends at a radius of $\sim$30\arcsec. \label{fig:modeln4559}}
\end{figure}

Finally, the last barred system discussed in this paper is NGC 4559,
an SABcd galaxy having a small bar roughly aligned with the major
axis ($PA_{bar} \sim$ 340\Deg\ vs $PA_{kin}=323$\Deg ). Numerical
simulations have been done for a bar parallel to the major axis and
are presented in the bottom panel of Figure \ref{fig:OH}. The small
velocity gradient seen in the PV diagram and in the rotation curve
is in agreement with the predicted behavior seen in the simulation.
The rotation velocities are artificially lowered by the radial
motions in the bar, resulting in an underestimation of the luminous
mass in the central regions. It is thus vital to model the bar in
order to determine the accurate mass distribution. The radial mass
distribution of the gas was analyzed between the 3 time steps to
verify that modifications of the inner part of the rotation curve
was not an effect of migration of the gas during the run.

Further bar modeling enables the bar length to be calculated.  Using
the Fourier moment analysis on azimuthal profiles derived using J
and K 2MASS images (see \citealt{1998AJ....116.2136A} and
\citealt{2000A&A...361..841A}), the deprojected value for the bar
semi--major axis is $30 \pm 5$\arcsec. For example, the tilted--ring
model for NGC 4559 shows that it is difficult to assess which radii
are affected by the bar (see Figure \ref{fig:modeln4559}).  However,
the \ha\ velocity field displays streaming motions outside the bar
as well, therefore the velocities are perturbed to at least a radius
of $\sim$40\arcsec. Deriving the form of the gravitational potential
directly from 2D kinematics and/or using numerical simulations of
the bar, are thus essential to perform in order to study properly
the mass distribution and solve the dark halo density profile
inconsistencies.

\section{Conclusions}
\label{conclusion}

We have presented in this paper the second and last part of the \ha\
kinematics follow--up survey of the Spitzer Infrared Nearby Galaxies
Survey (SINGS) sample. The goal of this kinematical follow up is to
better understand the role of baryons and of the dark/luminous
matter relation in star forming regions of galaxies.  The shape of
the velocity field in the central galactic regions, drawn through
its H$\alpha$ component, is indeed directly related to the baryonic
luminous disk and the star formation processes. The SINGS sample
will provide a unique opportunity to link the kinematics with
numerous observations and studies at other wavelengths.

The data have been obtained from high resolution Fabry--Perot
observations using the \FM\ camera and a L3CCD detector. The SINGS
sample of galaxies has been observed at the OmM 1.6m telescope, ESO
La Silla 3.6m telescope, CFHT 3.6m telescope, and WHT 4.2m
telescope.  The velocity fields were obtained by using a data
reduction pipeline written in IDL and the rotation curves were
computed by the \textit{rotcur} task from the {\it GIPSY} software.
When fitting the kinematical parameters, care was taken to avoid the
zones obviously affected by non--circular motions. However, we have
demonstrated that for barred systems, different bar characteristics
considerably modify the central velocity gradient of the computed
rotation curves in the pure circular rotation hypothesis. Therefore,
numerical modeling of barred galaxies is crucial in order to extract
rotation curves that are truly representative of the gravitational
potential and hence of the mass distribution in those galaxies. In
the meantime, the dark matter distribution of the SINGS galaxies,
using the rotation curves derived here, will be presented in a
forthcoming paper.

Not only will these observations provide the high spatial resolution
data needed for constraining the dark matter density profiles of
galaxies, but they will be helpful for studying these profiles as a
function of morphological type.  Furthermore, they will help to
delineate the role of gas kinematics in regulating the star
formation rate.  For instance, \cite{2002Ap&SS.281..101P} have
argued that the probability of collapse of molecular clouds leading
to star formation is greatly enhanced in slowly rotating gas discs
compared to high velocity rotation.  Moreover,
\cite{2003A&A...405...89C} and \cite{2007A&A...466..905F} have
suggested that the star forming inner and nuclear rings in the
nearby galaxies NGC 3627 and NGC 628 (respectively) are driven by a
rotating asymmetry because their location in the host disk are in
agreement with Lindblad resonances caused by a bar pattern speed.
Gas dynamical processes therefore are important in regulating the
star formation history of galaxies and the \ha\ kinematics presented
in this paper will help understanding the star formation processes.

\section*{Acknowledgments}

We would like to thank Jacques Boulesteix, Jean--Luc Gach, Philippe
Balard and Olivier Boissin for helping with the instrumentation and
part of the observations and the staff of the four Observatories,
where the data were obtained, for their continuing support. The
William Herschel Telescope is operated on the island of La Palma by
the Isaac Newton Group in the Spanish Observatorio del Roque de los
Muchachos of the Instituto de Astrofisica de Canarias. We
acknowledge support from the Natural Sciences and Engineering
Research Council of Canada and the Fonds Qu\'eb\'ecois de la
recherche sur la nature et les technologies. The Digitized Sky
Surveys (DSS images) were produced at the Space Telescope Science
Institute under U.S. Government grant NAG W--2166. The images of
these surveys are based on photographic data obtained using the
Oschin Schmidt Telescope on Palomar Mountain and the UK Schmidt
Telescope. The plates were processed into the present compressed
digital form with the permission of these institutions. The IR
images were obtained by the Spitzer Space Telescope, which is
operated by the Jet Propulsion Laboratory, California Institute of
Technology under a contract with NASA.

\appendix

\section[]{Description of the individual galaxies}
\hspace{5mm} \textbf{NGC 24}:  This galaxy is in the background of
the Sculptor Group of galaxies \citep{1988AJ.....95.1025P}. The
tilted--ring model fitted to the velocity field gives a mean
inclination of 75 $\pm$ 3\degr, which is consistent with the
photometric inclination but not with the \hi\ kinematical
inclination of 64 $\pm$ 3\degr\ derived in Chemin et al. (2006b). As
previously noticed by these authors, the low resolution \hi\ data
likely highly underestimates the inclination. A more detailed
description and analysis of our \ha\ data for this low surface
brightness galaxy will be presented in a forthcoming article (Chemin
et al. 2007, in prep.).

\textbf{NGC 337}: The spiral arms of this barred galaxy are easily
visible in the \ha\ integrated map.  The velocity field shows a
central Z--shape of the velocities indicating that a strong
disturbance caused by the bar. See more details in section
\ref{discussion}.

\textbf{NGC 855}: Not only bright \ha\ emission can be seen in this
dwarf elliptical galaxy, but CO emission has also been detected and
star formation is an ongoing process \citep{2007PASJ...59...61N}.
The \hi\ gas distribution is extended but the corresponding mass is
much smaller than for spirals \citep{1990ApJ...352..532W}.

\textbf{NGC 1097}: This large SBb galaxy exhibits a bright
circumnuclear ring with an associated steep velocity gradient.
Radial streaming motions from the nuclear ring to the galactic
center and a nuclear spiral are believed to be part of a mechanism
by which gas is fueled to the supermassive black hole
\citep{2006ApJ...641L..25F}. Those large streaming motions are most
likely responsible for the steep velocity gradient (up to nearly 300
\kms, see Fig. C1) seen toward the center. The PA of the bar is
aligned with both the galaxy's PA and companion, NGC 1097 A.  The
\hi\ gas is distributed fairly symmetrically in the prominent spiral
arms and in the bar.  Non--circular motions are found in and around
the bar as shown by the S--shape distortion of the \hi\ velocities
and the spiral arms display large streaming motions indicating the
presence of strong density waves \citep{1989ApJ...342...39O}.  CO
emission is detected mostly in the nuclear ring and the high
inferred \hmol\ mass can be explained by the secular action of the
bar enhanced by the interaction with NGC 1097 A
\citep{1988A&A...203...44G}.

\textbf{NGC 1291}: This early--type galaxy does not have enough
detected ionized gas for detailed kinematical mapping. In fact, only
nine \hii\ regions distributed in a ring--like structure can be seen
in the \ha\ monochromatic and RV maps.  The velocity range
corresponding to these \hii\ regions (835--855 \kms) is comparable
to what is seen in \hi\ observations, where the gas is distributed
in a bright ring surrounding a large central hole devoid of gas
\citep{1988A&A...204...39V}, typical of what is seen in the
bulge--dominated region of most early--type galaxies. The \hi\
velocity field of this face--on galaxy looks quite regular and the
associated rotation curve is essentially flat.

\textbf{NGC 1482}: The \ha\ distribution of this early--type galaxy
displays a bright central emission region and an eastern blob of
\ha\ emission.  The measured velocity dispersion of the lines in the
central region (some with $\sigma \ge 100$ \kms) is consistent with
the gas being ejected from the center.  \ha\ and [N{\scriptsize II}]
images by \cite{2002ApJ...565L..63V} revealed a galactic wind shaped
like an hourglass extending above and below the plane of the galaxy.
Both \ha\ and \hi\ observations show two bright blobs of emission
located on either side of the disc that are rotating about the
center of the galaxy \citep{2005MNRAS.356..998H}.  The CO global
profile presents a blueshifted component which is narrower and
weaker than the redshifted one \citep{1991ApJ...370..158S}.

\textbf{NGC 1512}: This strongly barred galaxy hosts two rings
inside the bar, a nuclear ring with strong \ha\ emission and large
velocity gradient and an inner ring located at the end of the bar
with weak emission compared to the nuclear ring.  The PA of the
nuclear ring is not perfectly aligned with those of the outer ring
and bar and the nuclear ring is off--centered with respect to the
inner ring. This \ha\ morphology is consistent with that presented
by \cite{1988ApJS...66..233B}. Their rotation model shows that the
nuclear ring must be expanding at a significant velocity (see also
\citealt{1981A&A....97...56L}).  This galaxy is likely to be
gravitationally interacting with its neighbor NGC 1510, as shown by
the perturbed \hi\ velocity field \citep{1979A&A....76..230H}. It is
worth also mentioning the presence of very extended tidal arms in
\ha\ and in UV, outside our FOV \citep{2006ApJS..165..307M}.

\textbf{NGC 1566}: The spiral arms of this grand--design galaxy are
well traced in the 3.6 $\mu$m and monochromatic images. Inside the
20\arcsec\ radius lies intense Doppler broadening typical of a
Seyfert I galaxy but too strong to be sampled by the free spectral
range of the FP interferometer used for our observations. Also,
there is an expanding bubble of gas located near
$\alpha=04^h19^m58^s$,
$\delta=-54^{\circ}55^{\prime}13^{\prime\prime} $ associated with
four strong \hii\ regions in the northwest arm and exhibiting a
strong velocity gradient.  These results are consistent with the
\ha\ observations made by \cite{1990ApJ...357..415P}. See
\cite{2004A&A...414..453A} for a dynamical analysis based on
long--slit spectroscopy.

\textbf{NGC 1705}: This peculiar galaxy shows a rotating disc that
is dominated by an intense ongoing starburst. Therefore, no
kinematical parameters could be extracted.  The double profiles and
multiple arcs of material visible in \ha\ suggest indeed violent
ejections of gas. In contrast, \hi\ synthesis observations revealed
a rotating disc in a dominant dark matter halo
\citep{1998MNRAS.300..705M}.

\textbf{Holmberg II}: The \ha\ content of this irregular galaxy
shows weak rotation and a lack of spatial coverage, therefore no
kinematical parameters or rotation curve could be extracted.
Besides, the neutral gas appears compressed on the southeast with a
large but faint component extending on the opposite side, indicating
ram pressure stripping from the intergalactic medium
\citep{2002AJ....123.1316B}.  The \hi\ velocity field displays a
weak but clear rotating disc pattern.  Like most of the dwarfs
\citep[see e.g.][]{1998ApJ...506..125C}, there is more luminous mass
in \hi\ than in stars, but contrary to most dwarfs where dark matter
dominates at nearly all radii, in HoII, dark matter dominates only
in the outer parts of the galaxy like in massive spirals.

\textbf{DDO 053}: This dwarf irregular galaxy has some diffuse \ha\
but since our observations have little spatial coverage, no
kinematical parameters are given for this galaxy.  The \hi\
distribution is compact and shows two peaks of emission with one
peak being associated with the brighter region visible in \ha\
\citep{2007ApJ...661..102W}.

\textbf{NGC 2841}: The velocity field of this fast--rotating galaxy
does not show any sign of perturbation.  The radial velocities
inside a 50'' radius are not mapped in this \ha\ study because of
Doppler broadening of the line typical of Seyfert 1 galaxies. Inside
the 50'' radius, \ha\ and [N{\scriptsize II}] images by
\cite{1999AJ....117.1725A} seem to suggest that the excitation of
the ionized gas is caused by shock--wave fronts.  Little CO is seen
in this galaxy \citep{2003ApJS..145..259H}.

\textbf{Holmberg I}: This dwarf galaxy has very little \ha\ emission
and the associated velocity field shows weak rotation.  The \hi\
morphology is characterized by one giant hole enclosed by a
ring--like structure.  The south side of the \hi\ ring is located so
that it encloses the \ha\ emission \citep{2007ApJ...661..102W}.

\textbf{NGC 3034 (M82)}: This starburst galaxy is seen nearly
edge--on and the opacity of the gas renders impossible the
extraction of the kinematical parameters.  The almost hourglass
shape of the \ha\ distribution and the presence of double profiles
above and below the disc indicates a remarkable outflow of gas
perpendicular to the plane of the galaxy.  Line--splitting is also
found in CO observations of the outflow and an estimated 10 percent
of the gas may be lost to the intergalactic medium
\citep{2002ApJ...580L..21W}. The FP data presented here agree
reasonably well with those of \cite{1998ApJ...493..129S},
particularly the strong redshifting and blueshifting seen north and
south of the disc respectively.  The differences in velocity between
the two FP data sets are mainly due to the high interference order
of our Fabry--Perot etalon which is designed for smaller radial
velocity range and higher spectral resolution. Fast ejection of
material is seen in the \ha\ velocity field, splitting the \ha\ line
by $\sim 300$ \kms.

\textbf{Holmberg IX}: There is no \ha\ emission originating from
this galaxy.  However, an expanding bubble of gas can be seen $\sim$
2\arcmin\ northeast of Holmberg IX.  This large supershell
($\sim250$ pc) is thought to be one of the most X--ray luminous
supernova remnants (SNRs) \citep{1995ApJ...446L..75M}.  The
position, velocity, and abundance similarities between Holmberg IX
and the superbubble suggest that they are related.

\textbf{NGC 3190}: This early--type galaxy has very little \ha.  A
close companion lies just 5\arcsec\ northwest and both galaxies are
in interaction.  No kinematical parameters are given for this
peculiar galaxy.

\textbf{IC 2574}: The \ha\ integrated map of this dwarf spiral
galaxy presents several supershells as well as large arc--like
shaped \hii\ complexes indicating gas compressed by powerful
star--forming events.  The kinematics yields a slowly rising
rotation curve until a value of V$_{rot} \sim 70 $ \kms\ is reached
\citep{2001AJ....121.1952B}. Mass modeling using the \hi\ rotation
curve from \cite{1994AJ....107..543M} yields a dark matter
contribution of 90\% which dominates at all radii. The \hi\
morphology is characterized by holes and several of the \ha\ regions
are located around the rims of the \hi\ holes. Finally,
\cite{2005ApJ...625..763L} have detected molecular gas in this
star--forming galaxy.

\textbf{NGC 3265}: This dwarf elliptical galaxy has a small rotating
\ha\ disc of radius $\sim 8$\arcsec\ in agreement with 21--cm line
observations.  The \hi\ rotation curve rises steeply until r =
20\arcsec\ then goes flat and the RV map shows presence of
non--circular motions or warping \citep{1987ApJ...314...57L}. Since,
for the few elliptical galaxies detected in \hi, the gas content is
not correlated with luminosity and the spatial distribution is not
reminiscent of that found in spiral galaxies, the origin of the gas
is thought to be external \citep{1987IAUS..127..145K}. Molecular gas
is detected in this elliptical galaxy and star formation is an
ongoing process in the centre \citep{1991ApJ...371..563G}.  Both CO
and \hi\ profiles have half--intensity widths of $\sim 180$ \kms.

\textbf{Mrk 33}: The \ha\ emission of this dwarf starburst galaxy,
also known as Haro 2, reveals at least two star--forming regions in
the galactic centre and the overall kinematics resemble a luminous
oblong expanding shell.   A more in--depth observation by
\cite{2000A&A...359..493M} has shown three bright \ha\ knots and
faint filamentary structures.  The presence of an extended X--ray
emitting region lying within the supershell suggests a superbubble
caused by a starburst--driven outflow
(\citealt{2001astro.ph..6475S}, \citealt{1997A&A...326..929L},
\citealt{1995A&A...301...18L}).  No kinematical parameters are given
for this galaxy.

\textbf{NGC 3351 (M95)}: The \ha\ distribution in this barred galaxy
shows a luminous centre with an \ha\ ring and the corresponding
velocity field presents evidence of strong non--circular motions in
the centre.  Kinematics inside this region reveal that the stellar
bar is radially driving gas towards the star forming \ha\ ring at V
= 25 \kms \citep{2007MNRAS.378..163H}, from elongated x1 orbits to
the x2 orbits closer to the bar. Furthermore, the detected excess in
the central ISM content of the large--scale bar implies that the
latter is transporting gas to smaller radii
\citep{2006ApJ...652.1112R}. Interestingly, CO observations have
shown a small molecular bar aligned perpendicular to the large scale
stellar bar (\citealt{2003ApJS..145..259H},
\citealt{1992AJ....103..784D}).  The earlier paper argues that a
resonance caused by the stellar bar lies in the vicinity of the \ha\
ring and nuclear molecular gas bar.  The velocity gradient in the
central regions is steep in both CO and \ha\ data.  There is also an
inner ring at the end of the bar where the velocities are regular
and not perturbed.  Another \ha\ kinematical analysis has been done
for this galaxy by \cite{1988ApJS...66..233B}.  Finally, two nuclear
spiral arms residing inside the circumnuclear ring can be seen in
{\it Spitzer} IRAC 3.6$\umu$m data.

\textbf{NGC 3627 (M66)}: This member of the Leo triplet shows a
highly perturbed \ha\ morphology and velocity field.  Streaming
motions can be seen along the spiral arms and a warp can be fitted
on the southern, receding side of the disc.  The \ha\ kinematics is
consistent with CO and \hi\ data obtained by
\cite{2003ApJS..145..259H} and \cite{2002ApJ...574..126R}
respectively.  A molecular and \ha\ ring has been found lying at the
position of the ultra--harmonic resonance induced by the bar pattern
speed \citep{2003A&A...405...89C}.  The proposed explanation for the
asymmetric and perturbed gas morphology is a tidal interaction with
NGC 3628 (\cite{2005A&A...429..825A}; see also
\cite{1993ApJ...418..100Z}).

\textbf{NGC 3773}: Visible in both \ha\ and \hi\ data, this
early--type galaxy displays gas concentration in its centre.  There
is no detectable rotation and the global profile is very narrow,
both for the ionized and neutral gas \citep{1987AJ.....94..883B}.

\textbf{NGC 4254(M99)}: The kinematical analysis of this grand
design spiral galaxy has already been presented in the Virgo galaxy
cluster sample of \cite{2006MNRAS.366..812C}. The \ha\ velocity
field presents significant streaming motions along the spiral
structure.  Recent observations by \cite{2007ApJ...665L..19H}
reported an \hi\ tail extending $\sim$ 250 kpc to the North of NGC
4254, perhaps the result of galaxy harassment as the galaxy enters
the Virgo cluster.

\textbf{NGC 4450}: The Fabry--Perot data of this anemic galaxy have
been presented in \cite{2006MNRAS.366..812C}. The \ha\ distribution
is very clumpy and the velocity field perturbed in the innermost
regions of the galaxy.  A new tilted--ring model is fitted to the
velocity field using rings of 4\arcsec\ width instead of 2\arcsec\
in our previous analysis. This allows one to derive a new major axis
position angle of 353$\pm$5\degr\ with a smaller error bar than in
\cite{2006AJ....132.2527C}. The values remain in agreement within
the errors.

\textbf{NGC 4559}: The \ha\ velocity field of this late--type galaxy
shows the presence of streaming motions in its centre and is
somewhat patchy far from the centre.  The \hi\ disc, extending
further out than the optical disc, is warped and lopsided in both
distribution and kinematics \citep{2005A&A...439..947B}.  Their \hi\
PV diagram revealed the presence of a thick \hi\ layer rotating
25--50 \kms\ more slowly than the value for the thin disc.

\textbf{NGC 4594 (M104)}: This early--type galaxy is better known as
the Sombrero galaxy. The \ha\ observations display only one diffuse
ionized region west of the nearly edge--on disc and the
corresponding radial velocities are in agreement with the galaxy's
systemic velocity.  Stellar kinematics reveal for the rotation curve
a rapid rise followed by a decrease which can be explained by the
prominent galactic bulge \citep{1993MNRAS.263.1049C}.  Mass modeling
using globular cluster kinematics has shown that the $(M/L)$ value
increases with radius, hence that M104 has a dark matter halo
\citep{1997MNRAS.284..376B}.  Another study of the dark matter
distribution has been done by \cite{2006MNRAS.371.1269T}.

\textbf{NGC 4631}: This galaxy is seen edge--on and its inclination
renders impossible the extraction of the kinematical parameters. The
interaction with the nearby dwarf galaxy is thought to be
responsible for the warp seen on the southeast side of the \ha\
velocity field.  Our \ha\ results are consistent with those of
\cite{1996A&A...313..439G}.  As for the \hi\ data, the results
obtained by \cite{1994A&A...285..833R} show that the maximum
rotation velocity is 140 \kms and that the \hi\ disc can be modeled
as having two distinct velocity components.  CO observations have
demonstrated that the sources of these distinct velocities are the
spiral arms within the gas disc \citep{1990PASJ...42..745S}.

\textbf{NGC 4736 (M94)}: This ringed galaxy exhibits a bright inner
ring with a nearly constant rotational velocity of about 195 \kms.
The dispersion velocities are typical of those seen in \hii\
regions.  The location of this \ha\ ring is consistent with
resonances caused by the bar and by a larger oval distortion.  CO
observations reveal tightly wound spiral arms separated by a small
nuclear bar that is perpendicular to the major axis PA of the galaxy
\citep{2003ApJS..145..259H} as well as gas moving on elliptical
orbits around the nuclear bar.  Modeling of \ha\ and CO kinematics
shows an inflow of material near the ends of the nuclear bar, an
outflow between the bar and the ring, and an inflow of gas just
outside the ring (\citealt{2004AJ....127...58M},
\citealt{2000ApJ...540..771W}).  The slightly declining \ha\
rotation curve presented in this paper is consistent with CO and
\hi\ observations, but not in the central regions.

\textbf{DDO 154}: This dwarf irregular galaxy has only a few
discrete \hii\ regions and larger diffuse regions.  The \ha\
velocity field shows large--scale but slow rotation ($\sim 30$ \kms)
that can be measured to just below 2\arcmin. The 21--cm \hi\
observations present a regular but extended gas disc having a slight
warp at the southwest end (\citealt{1993AJ....106...39H},
\citealt{1989ApJ...347..760C}).  Mass modeling of this galaxy
reveals that more than 90\% of the mass is found in the dark
component (\citealt{1998ApJ...506..125C},
\citealt{1988ApJ...332L..33C}).

\textbf{NGC 4826 (M64)}: The \ha\ distribution of this galaxy,
confined to the inner disc, is more extended on the northwest
receding side than on the approaching side, which is consistent with
CO emission \citep{2003ApJS..145..259H}.  The \hi\ gas, on the other
end, extends far beyond the optical radius $R_{25}$ and is also
distributed asymmetrically.  The \hi\ velocity field reveals a
counter--rotating gaseous disc outside the 2\arcmin\ diameter
\citep{1992Natur.360..442B}.  A detailed kinematical study, based on
\ha\ long--slit spectroscopy, reveals a transition region near the
prominent dust lane (50\Sec$\le$r$\le$70\Sec) where the gas motion
is undergoing a change in its orbital direction, from prograde in
the inner disc to retrograde in the outer disc
\citep{1994AJ....107..173R}.  This situation, symptomatic of a
merger, has been widely seen in elliptical galaxies, but rarely in
disc galaxies.

\textbf{DDO 165}: The \ha\ map of this dwarf galaxy presents only
four \hii\ regions that may not be all physically related to the
galaxy.  The \hi\ distribution is extended and is somewhat arc--like
shaped \citep{2007ApJ...661..102W}.

\textbf{NGC 5033}: The \ha\ velocity field displays a strong
velocity gradient ($\sim 100$ \kms per kpc) in the centre of this
galaxy, agreeing with CO emission \citep{2003ApJS..145..259H}.  The
broad profiles found near the centre of the galaxy are
characteristic of a Seyfert galaxy.  \oiii\ and \hb\ observations
show respectively an asymmetric morphology and an off--centered
active nucleus which can be understood in terms of a past merger
\citep{2005A&A...433...79M}.  The \hi\ large--scale velocity field
is well ordered and streaming motions can be seen near the spiral
structures (see \citealt{1997MNRAS.290...15T} and
\citealt{1981AJ.....86.1791B}).  The \ha\ and \hi\ rotation curves
present a constant rotational velocity of $\sim 215$ \kms\ and a
warp developing beyond the optical disc.

\textbf{NGC 5408}: This irregular galaxy has a highly asymmetric
\ha\ morphology with a bright nuclear region situated at one end of
an elongated optical distribution.  The FP observations are in
agreement with those of \cite{1972ApJ...175..329B}.  The absence of
large--scale motions is the reason why no kinematical parameters are
given for this galaxy.

\textbf{NGC 5474}: The \ha\ distribution of this peculiar galaxy is
composed of several \hii\ regions and clumpy diffuse ionized gas
which are rotating weakly. The \hi\ disc also displays a clumpy but
fairly symmetric distribution.  The \hi\ velocity field exhibits
normal differential rotation but warping near the edge of the \hi\
disc can be seen (\citealt{1994AJ....108.1638R}, see also
\citealt{1979A&A....78...82V}).

\textbf{NGC 6822}: This nearby irregular galaxy does not exhibit
strong rotation within the FOV of our observations, while some is
seen in the \hi\ data  \citep{2000ApJ...537L..95D}. The \ha\
emission is composed of \hii\ regions and low density regions of
diffuse ionized gas. The \hi\ distribution is far more extended and
is dominated by holes and clouds.

\textbf{NGC 7552}: This SBab galaxy is completely dominated by its
bar, making the extraction of kinematical parameters impossible. The
velocity field shows a central S--shape disturbance indicating
non--circular motions caused by the bar. A steep velocity gradient
($\sim 150$ \kms\ per kpc) is present in the nuclear region as well
as a bright \ha\ centre representative of a LINER galaxy.

\textbf{NGC 7793}: The close--up view of this galaxy displays
several \hii\ regions and the corresponding \ha\ velocity field
exhibits an s--shape distortion of the velocities along the minor
axis. A kinematical analysis of the fairly regular \hi\ velocity
field reveals a declining rotation curve
\citep{1990AJ....100..394C}. Deep FP observations of the diffuse gas
have shown an extended \ha\ disc that is detected all the way to the
edge of the \hi\ disc.

\section{Figures}

Figures were removed for the astro-ph version of this paper due to file size constraints.  You can access them online at http://www.astro.umontreal.ca/fantomm/singsII/.

\section{Rotation curves}
\label{app:rc}

\begin{figure*}
\centering
\vspace{1cm}
{
  \includegraphics[width=0.45\textwidth]{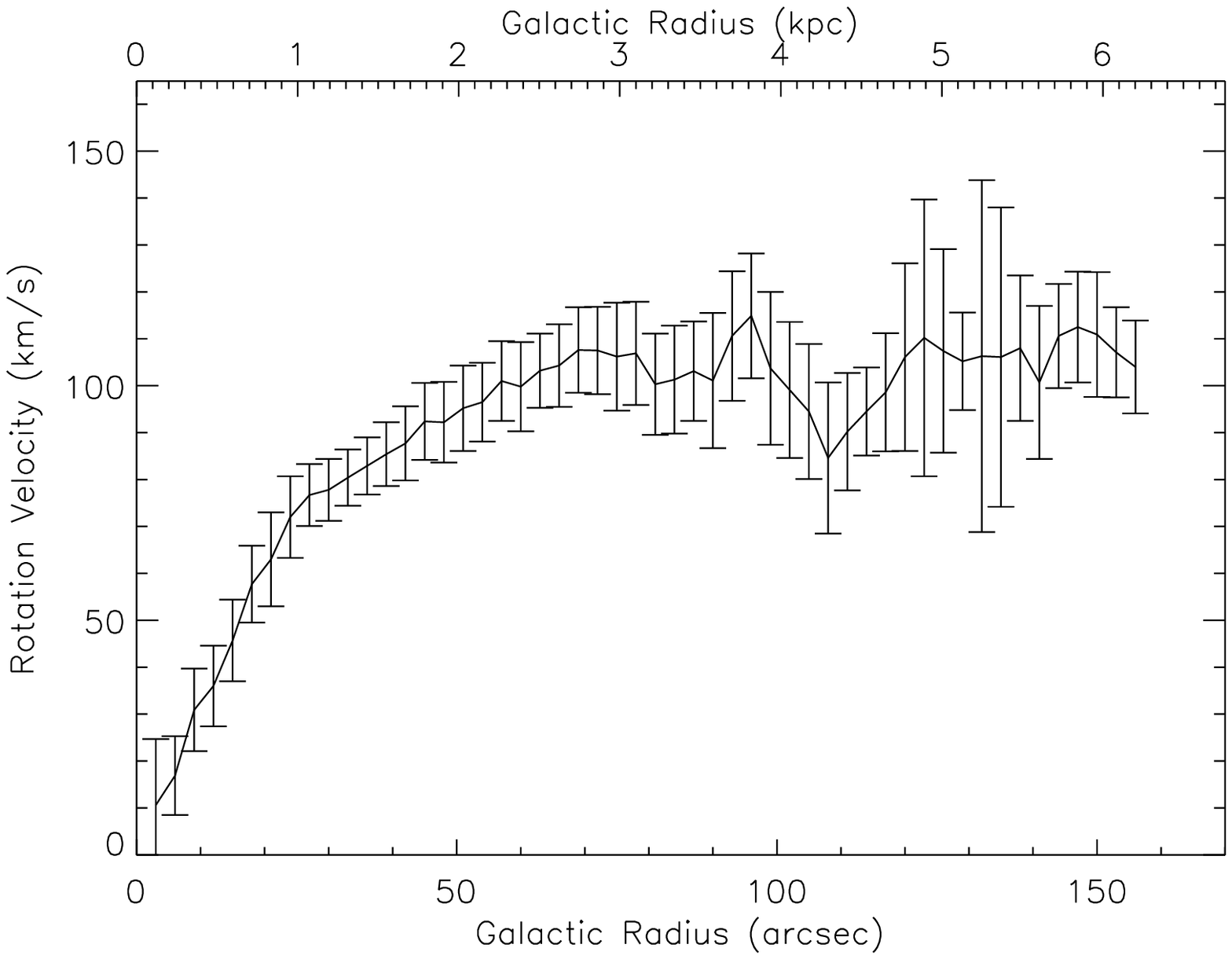}
}
\hspace{1cm}
{
   \includegraphics[width=0.45\textwidth]{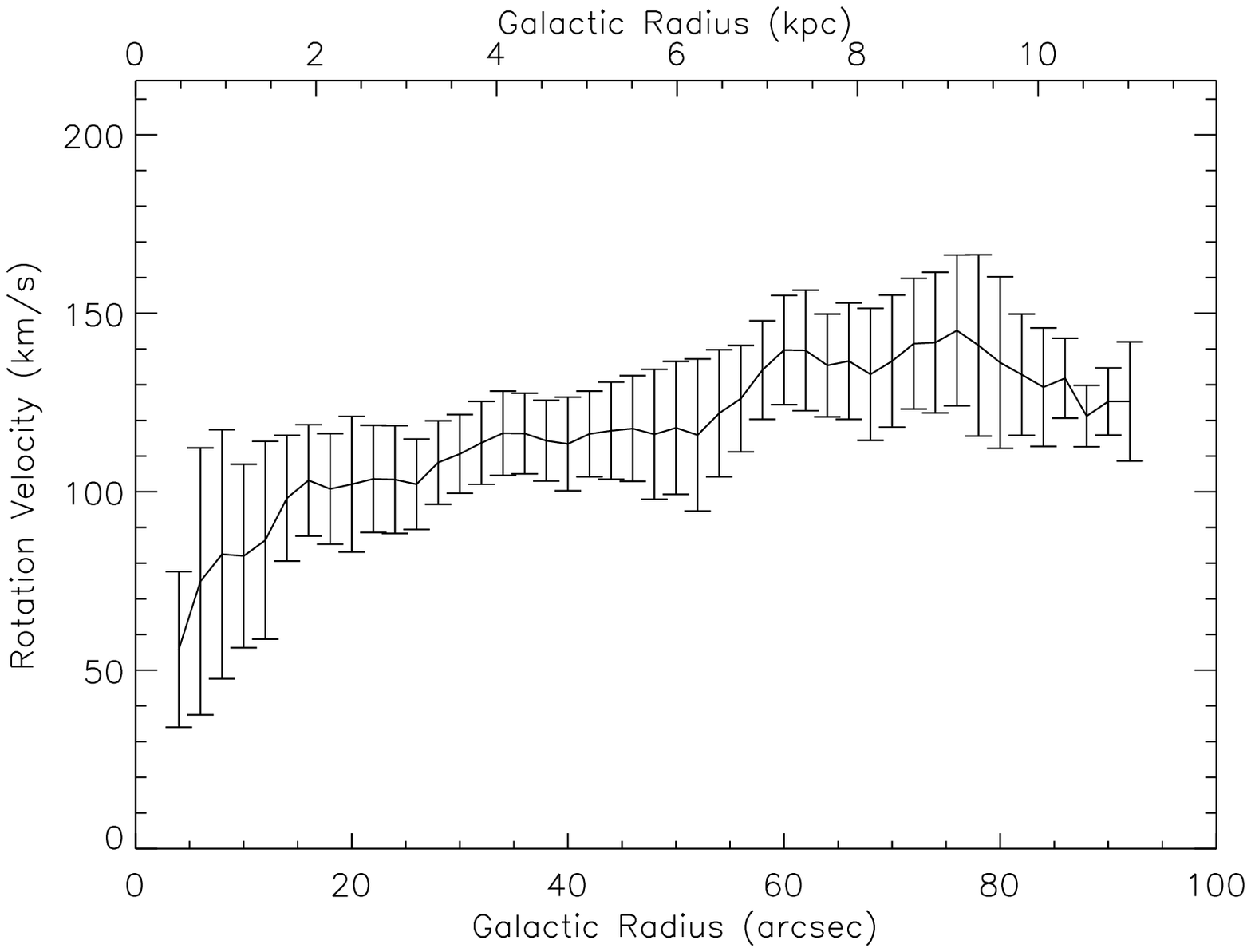}
}
\vspace{1cm}
{
 \includegraphics[width=0.45\textwidth]{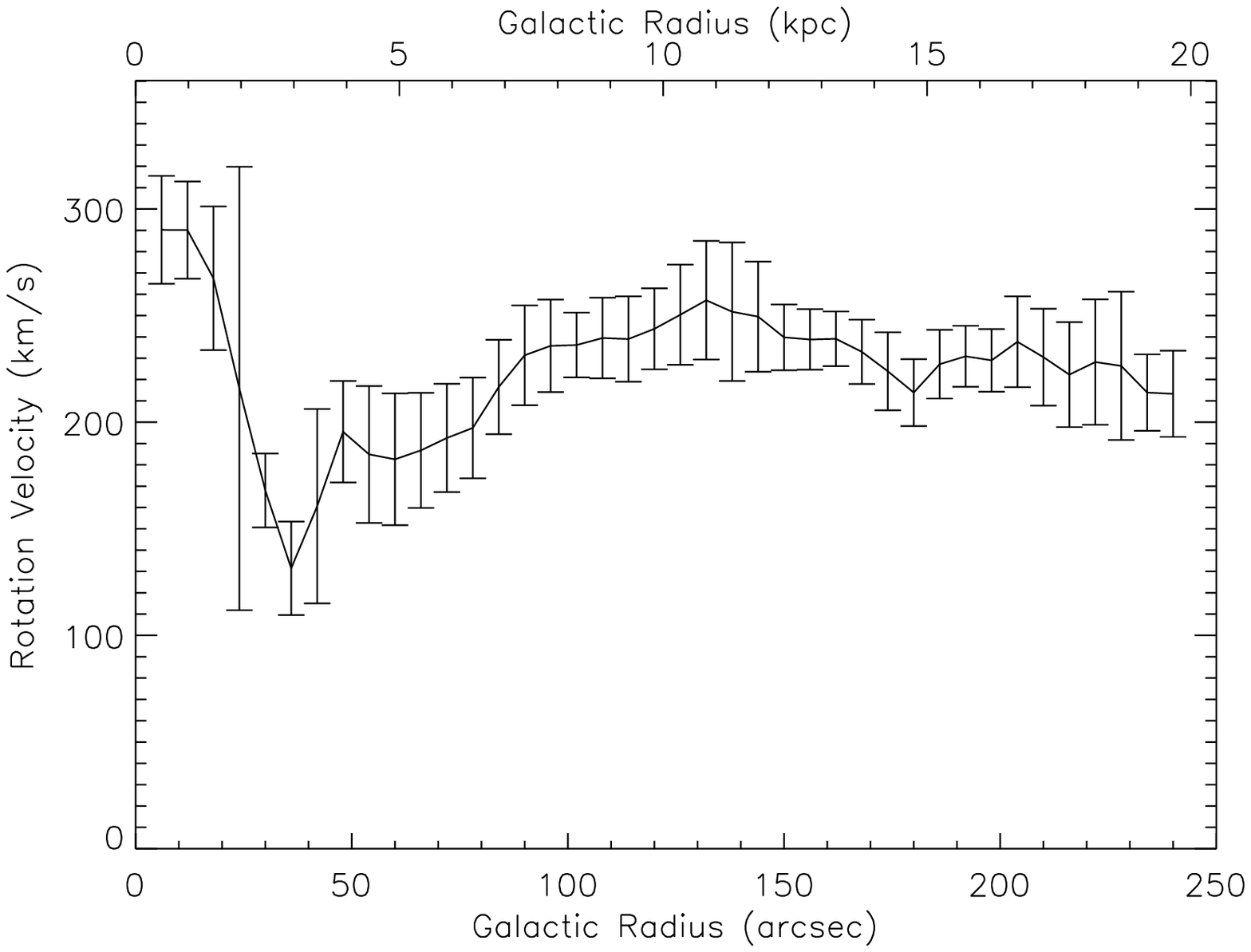}
}
\hspace{1cm}
{
     \includegraphics[width=0.45\textwidth]{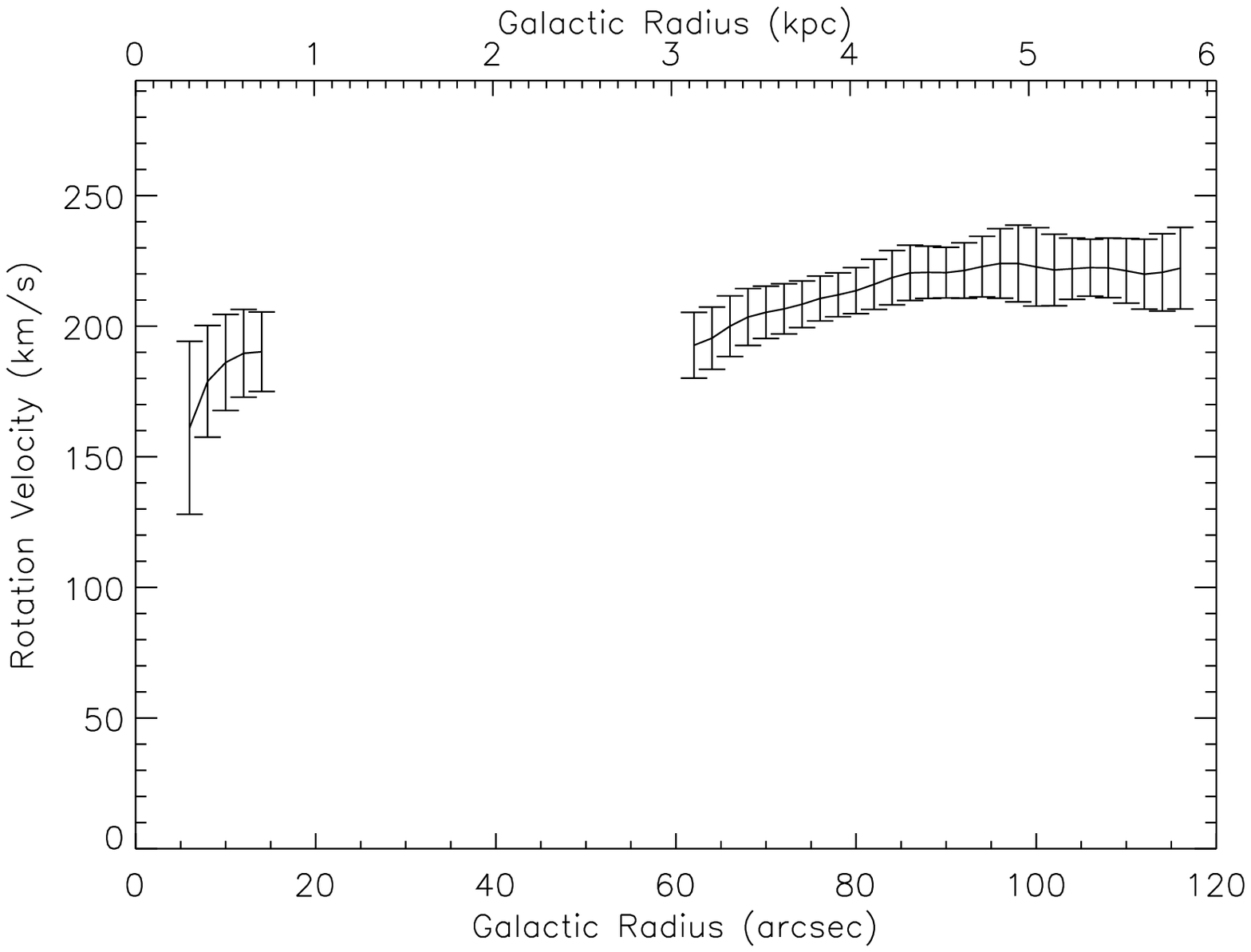}
}
\vspace{1cm}
{
\includegraphics[width=0.45\textwidth]{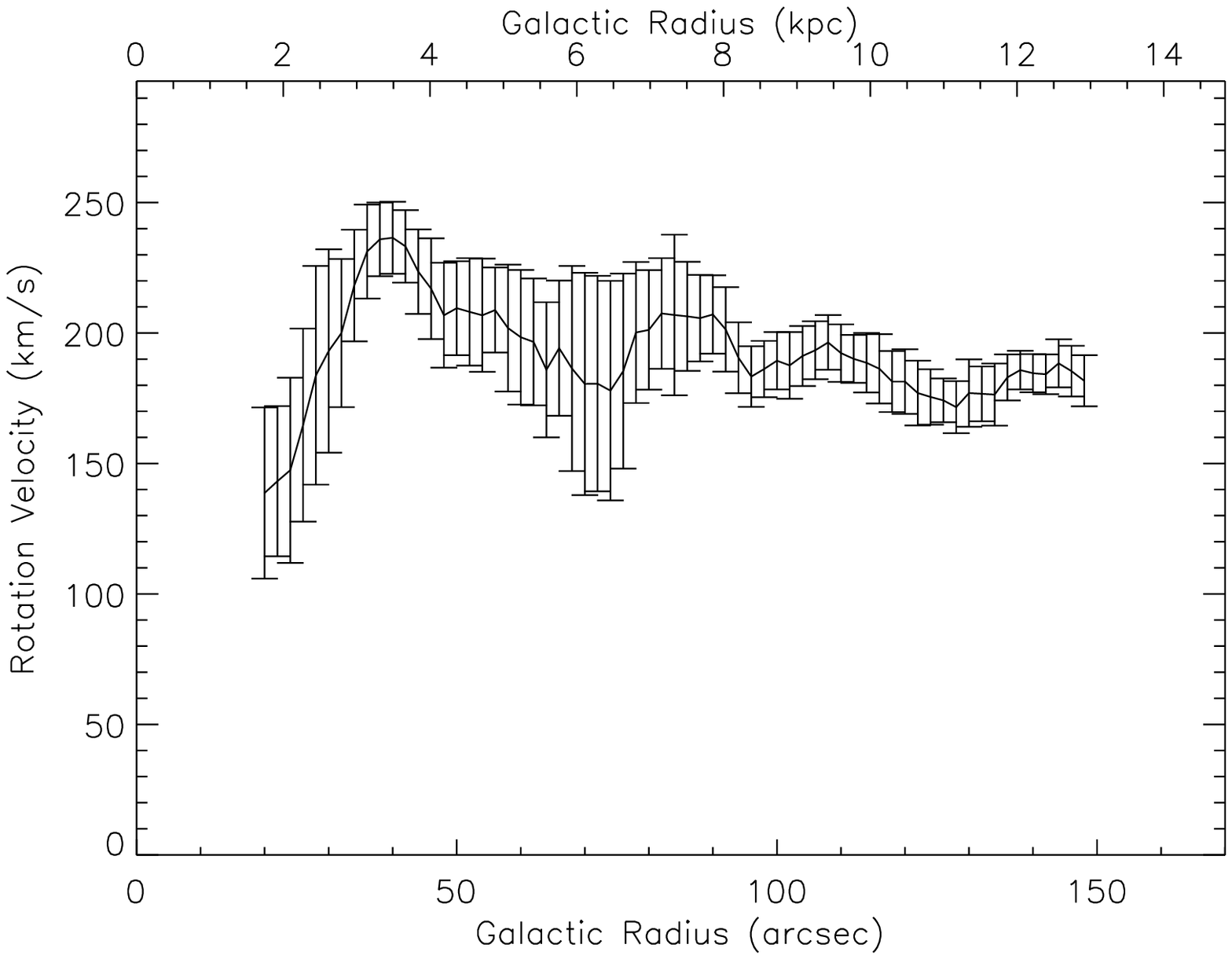}
}
\hspace{1cm}
{
    \includegraphics[width=0.45\textwidth]{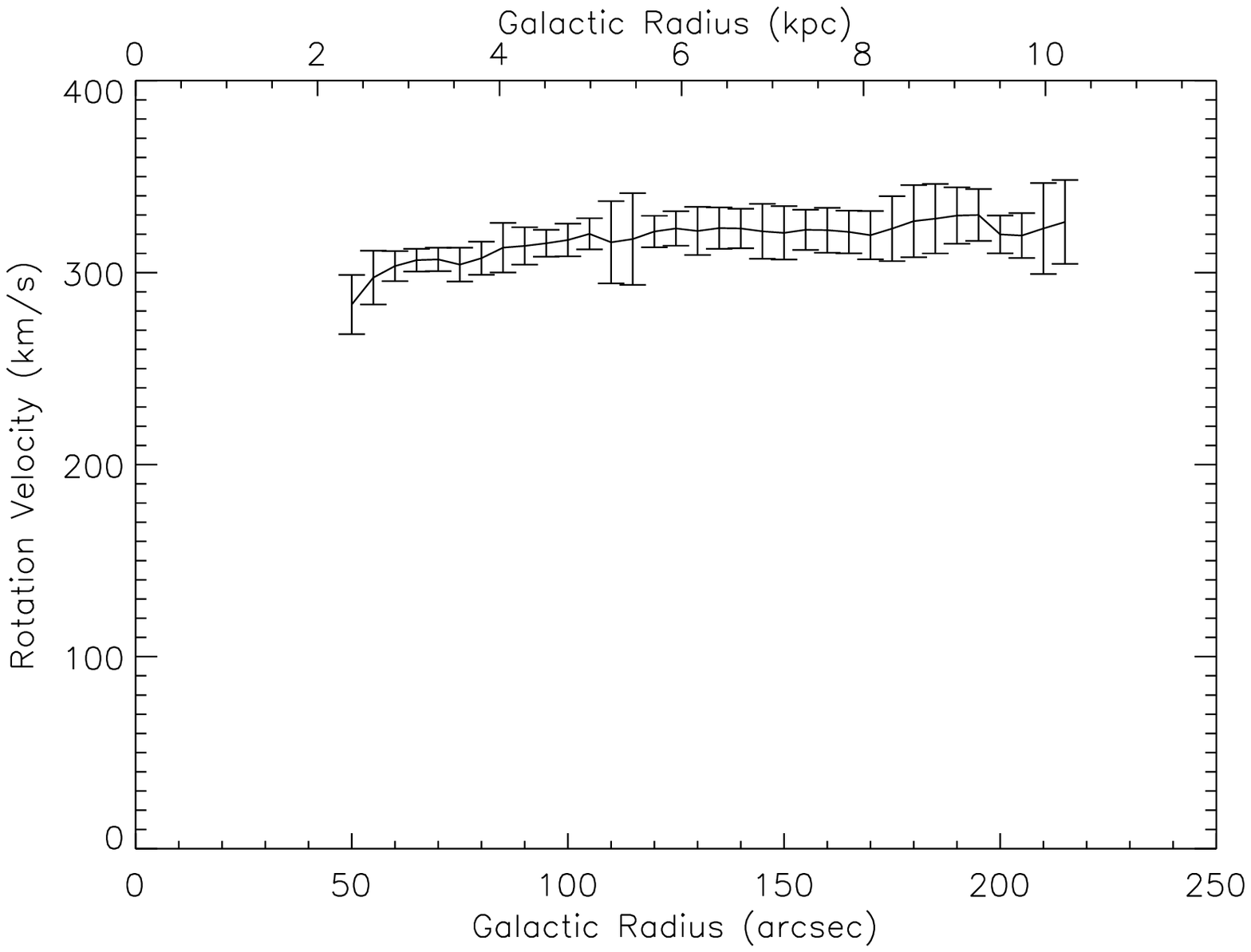}
} \vspace{1cm} \caption{Rotation curves for the galaxies NGC 24
(top--left: PA=226\Deg, i=75\Deg), NGC 337 (top--right: PA=121\Deg,
i=52\Deg), NGC 1097 (middle--left: PA=133\Deg, i=55\Deg), NGC 1512
(middle--right: PA=260\Deg, i=35\Deg), NGC 1566 (bottom--left:
PA=214\Deg, I=32\Deg), and NGC 2841 (bottom--right: PA=150\Deg,
i=70\Deg).}
\end{figure*}

\begin{figure*}
\centering
\vspace{1cm}
{
    \includegraphics[width=0.45\textwidth]{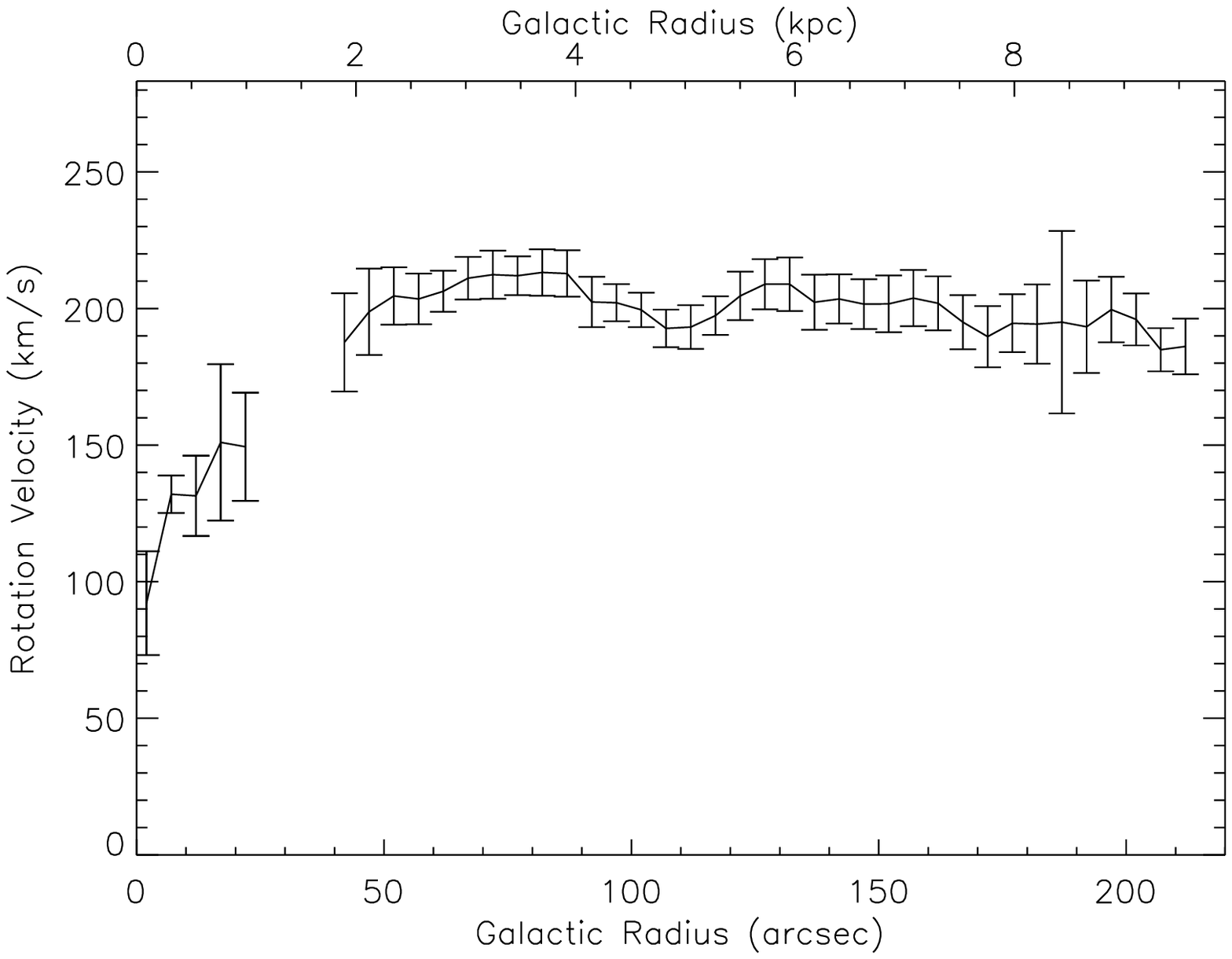}
}
\hspace{1cm}
{
    \includegraphics[width=0.45\textwidth]{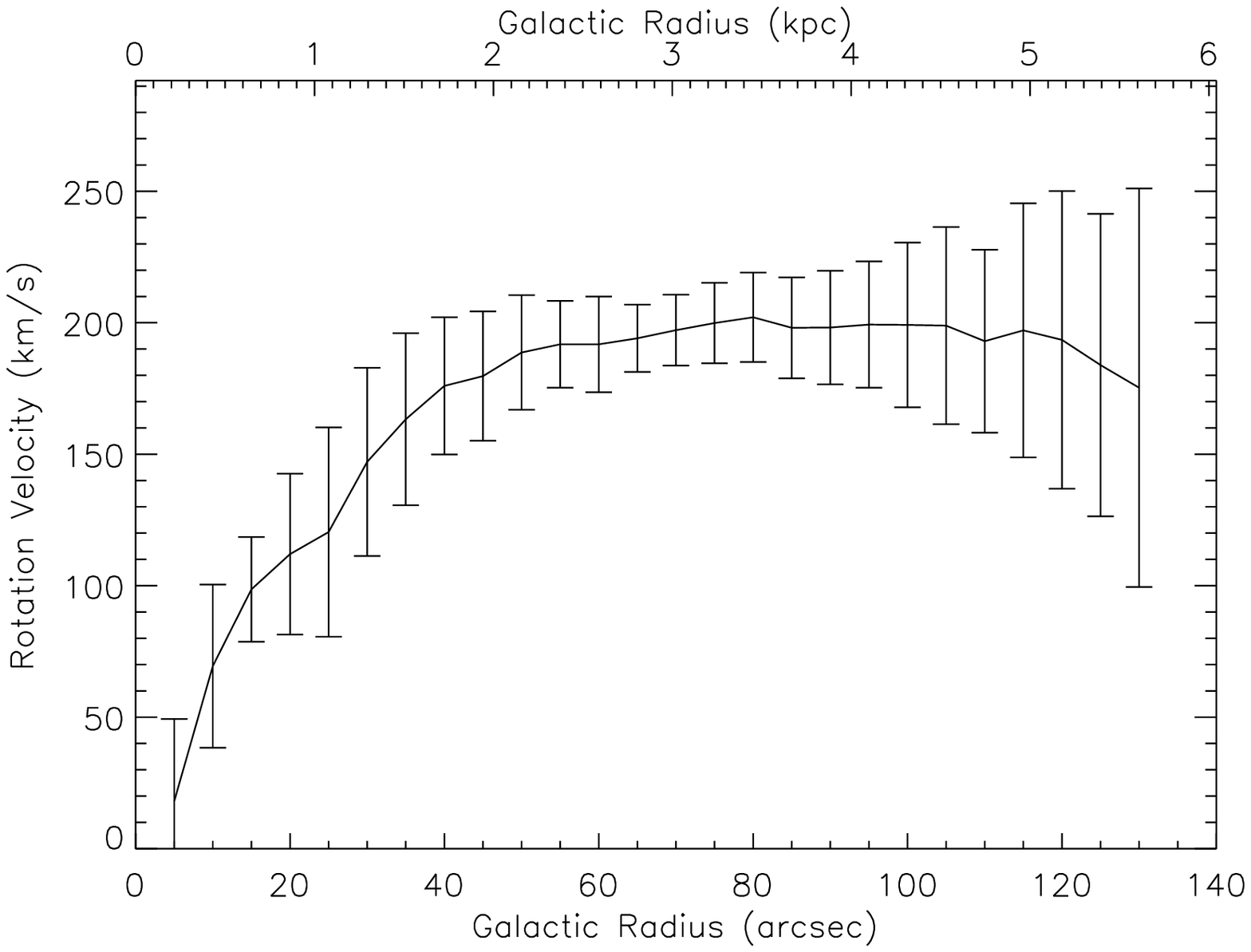}
}
\vspace{1cm}
{
   \includegraphics[width=0.45\textwidth]{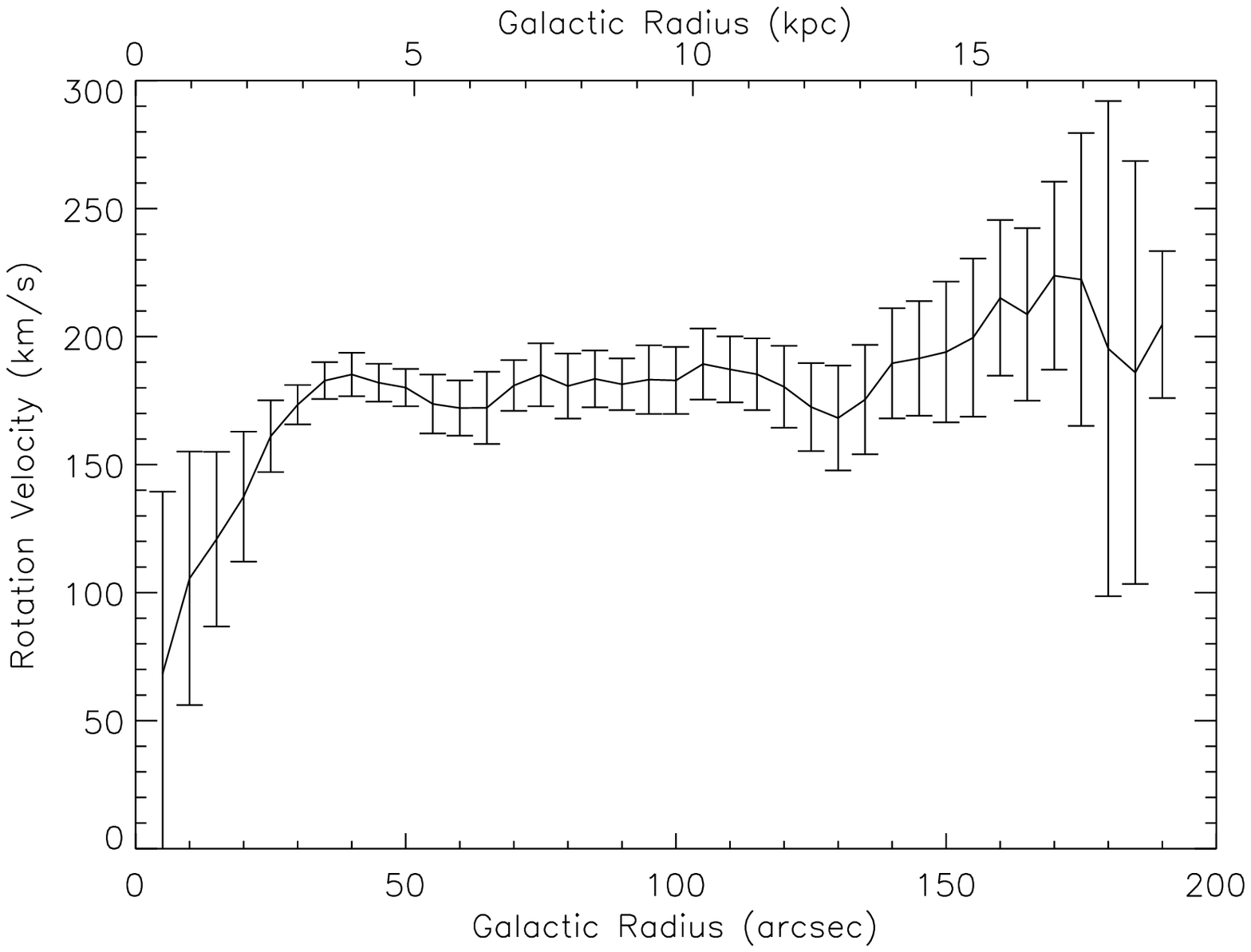}
}
\hspace{1cm}
{
    \includegraphics[width=0.45\textwidth]{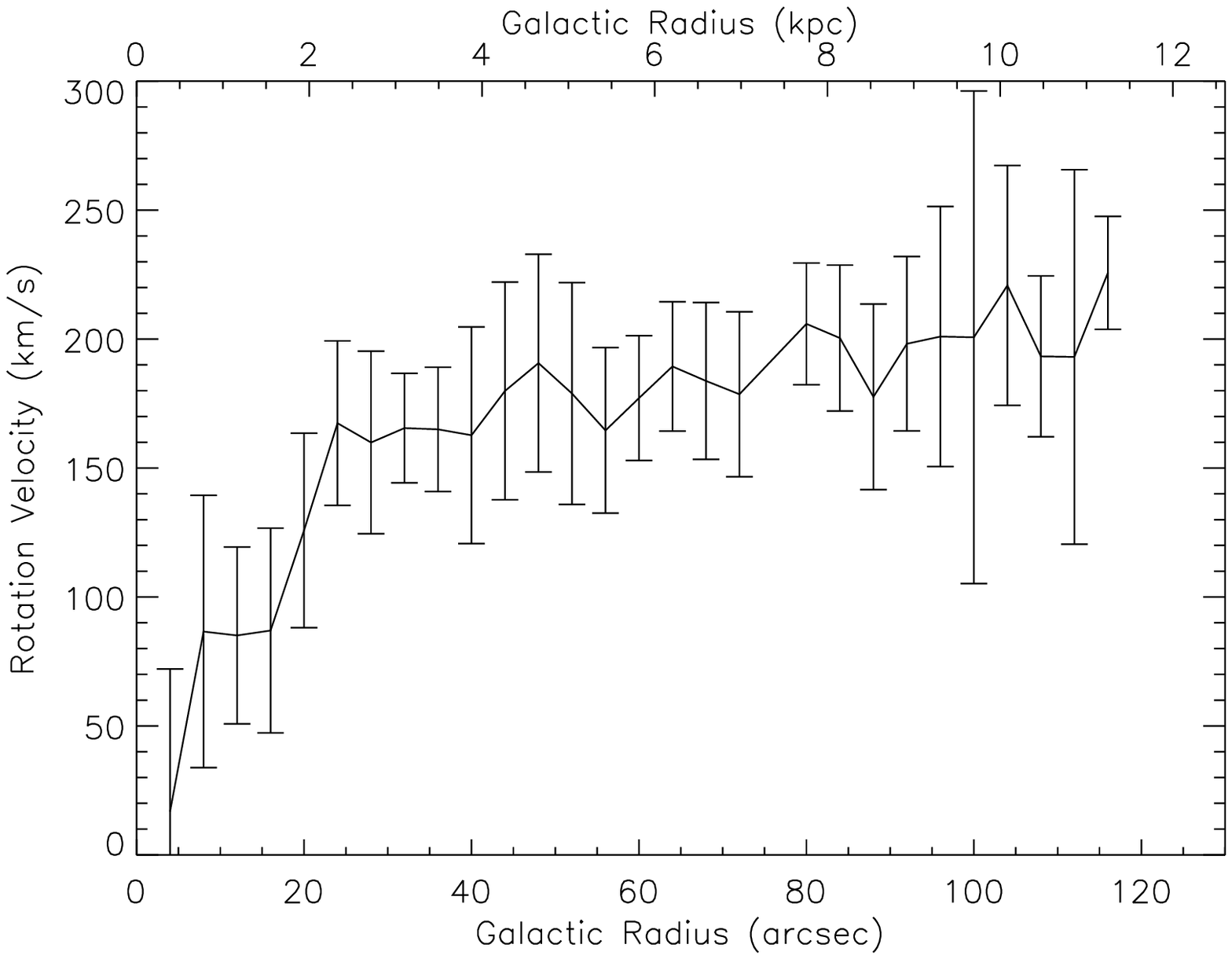}
}
\vspace{1cm}
{
    \includegraphics[width=0.45\textwidth]{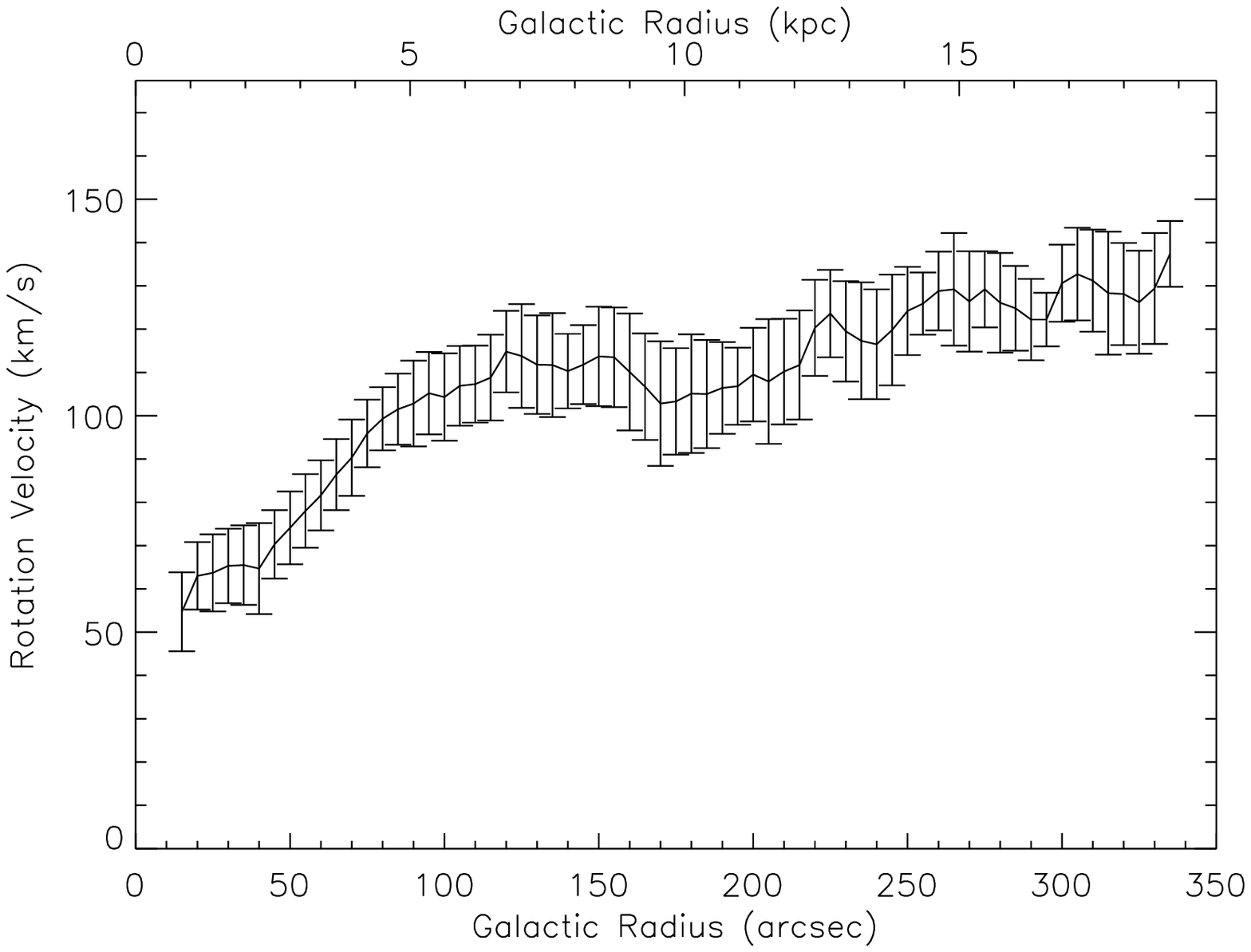}
}
\hspace{1cm}
{
    \includegraphics[width=0.45\textwidth]{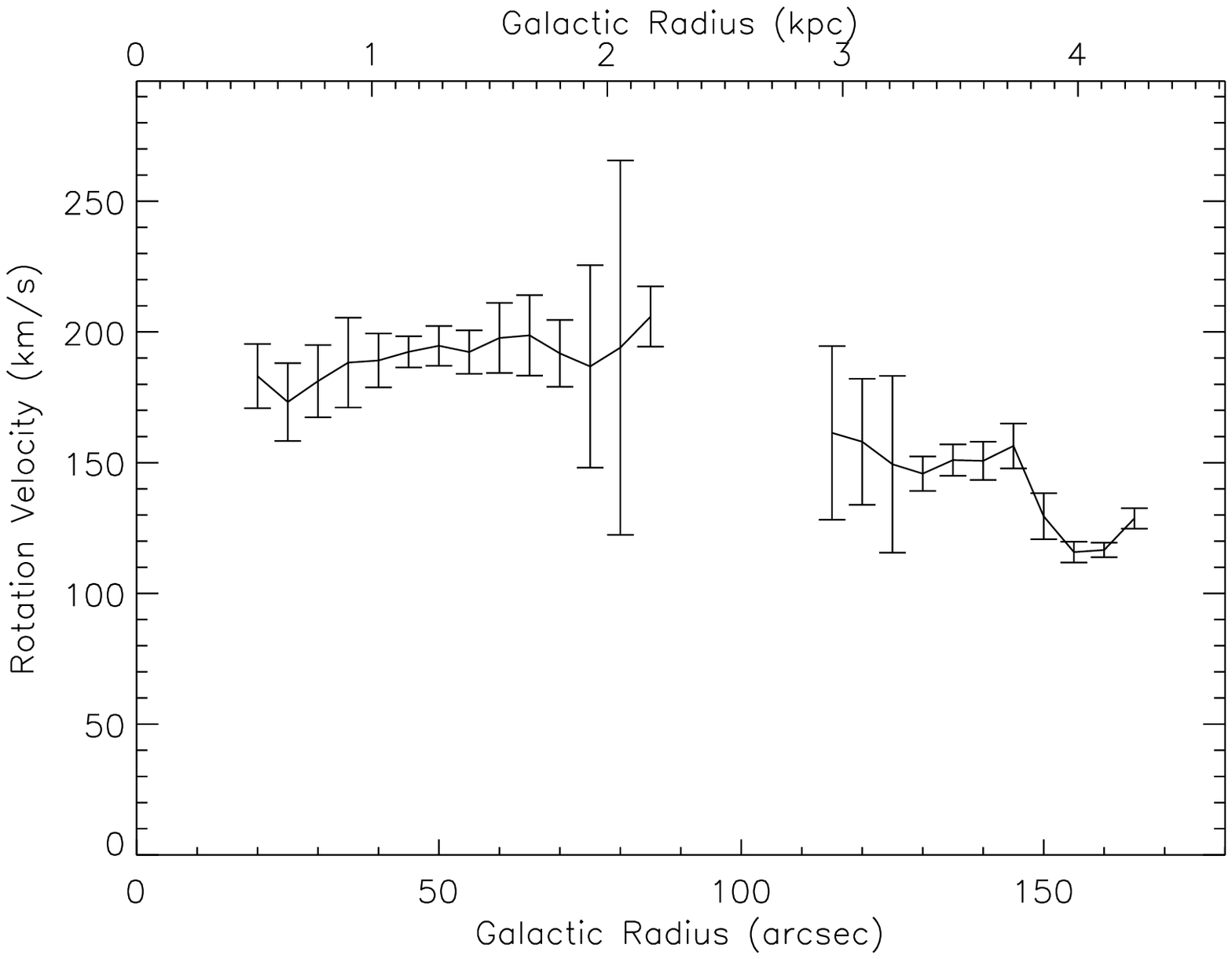}
} \vspace{1cm} \caption{Rotation curves for the galaxies NGC 3351
(top--left: PA=193\Deg, i=41\Deg), NGC 3627 (top--right: PA=173\Deg,
i=65\Deg), NGC 4254 (middle--left: PA=69\Deg, i=31\Deg), NGC 4450
(middle--right: PA=353\Deg, i=43\Deg), NGC 4559 (bottom--left: PA=
323\Deg, i=68\Deg), and NGC 4736 (bottom--right: PA= 292\Deg,
i=36\Deg).}
\end{figure*}

\begin{figure*}
\centering
\vspace{1cm}
{
   \includegraphics[width=0.45\textwidth]{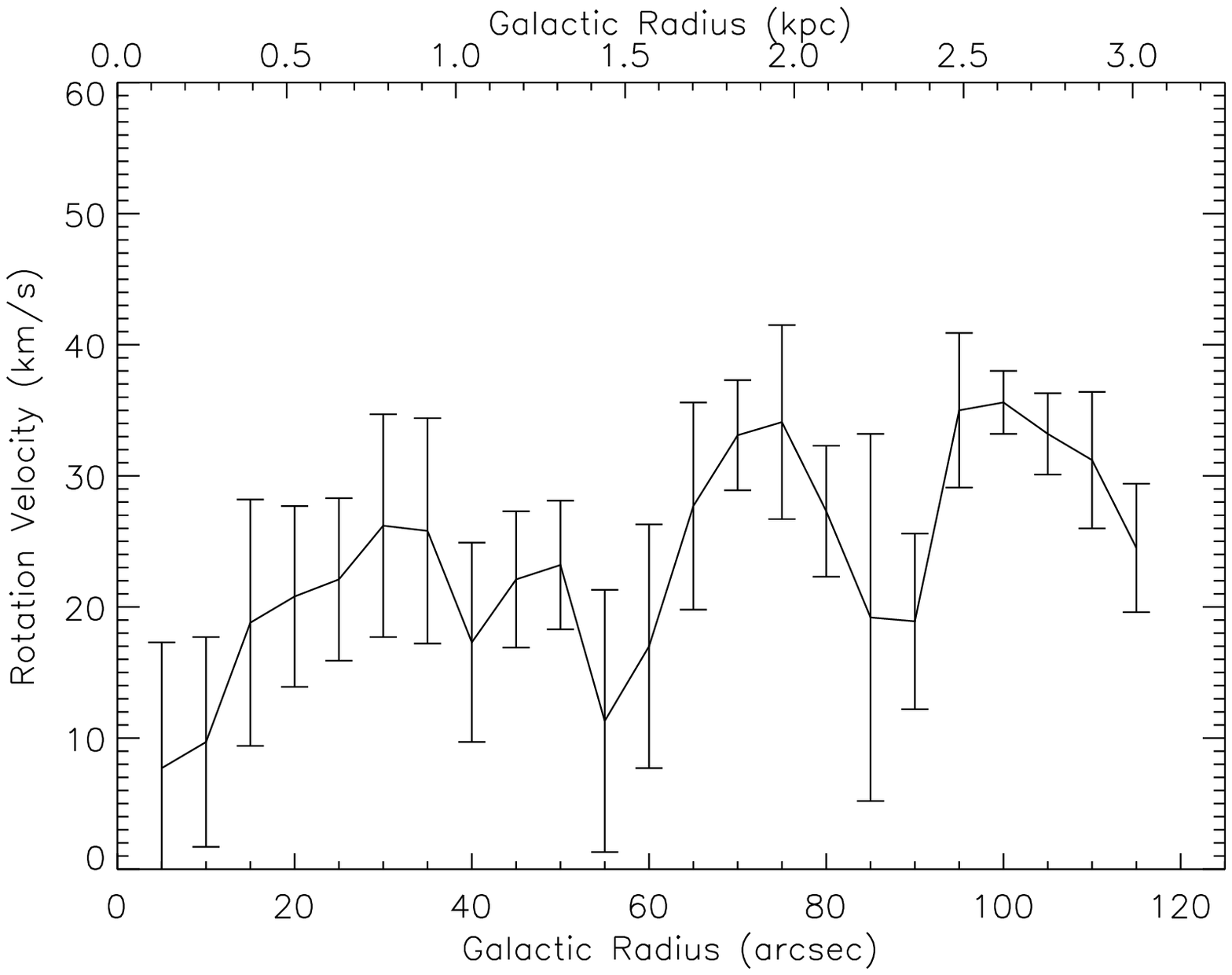}
}
\hspace{1cm}
{
    \includegraphics[width=0.45\textwidth]{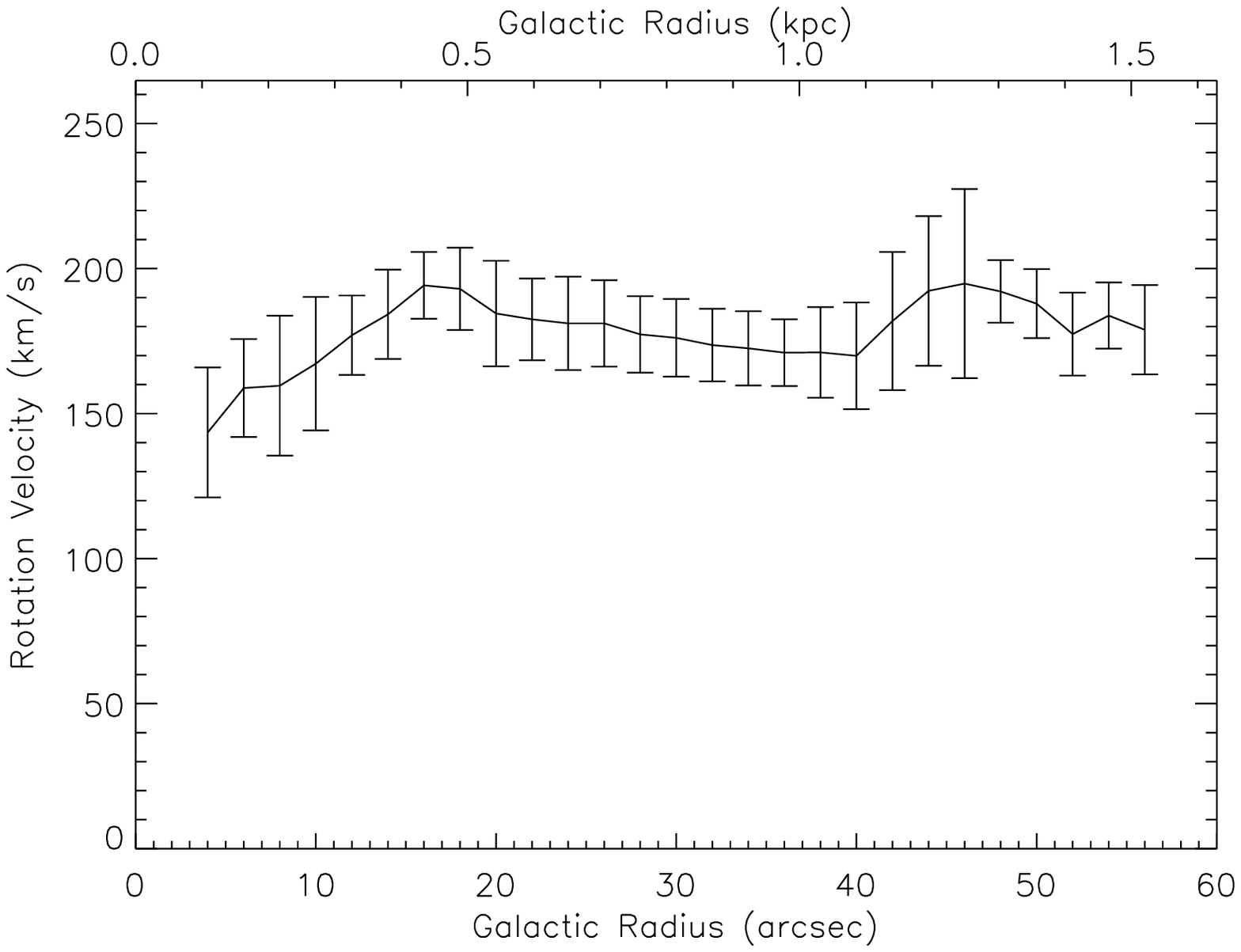}
}
\vspace{1cm}
{
    \includegraphics[width=0.45\textwidth]{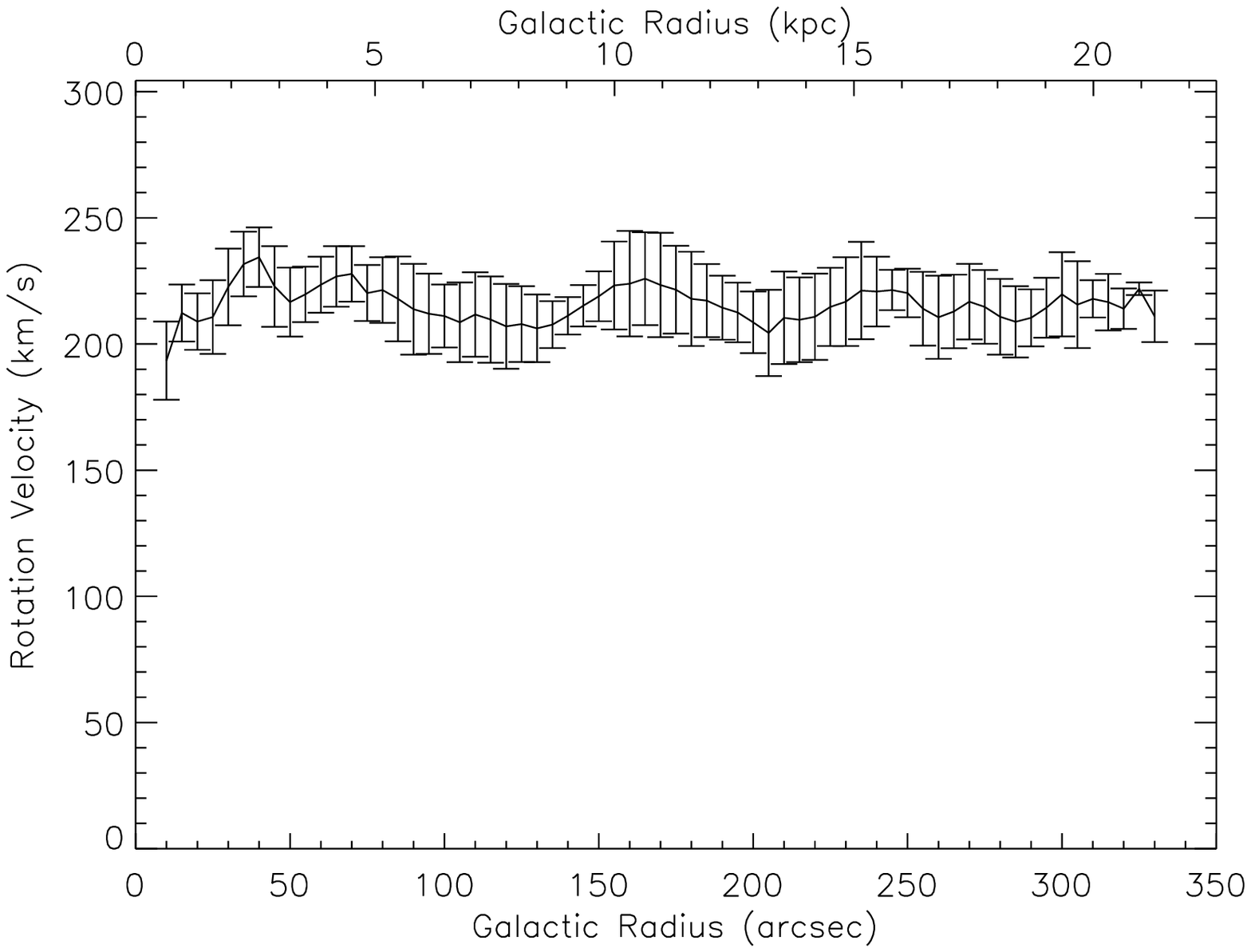}
} \hspace{1cm} {
   \includegraphics[width=0.45\textwidth]{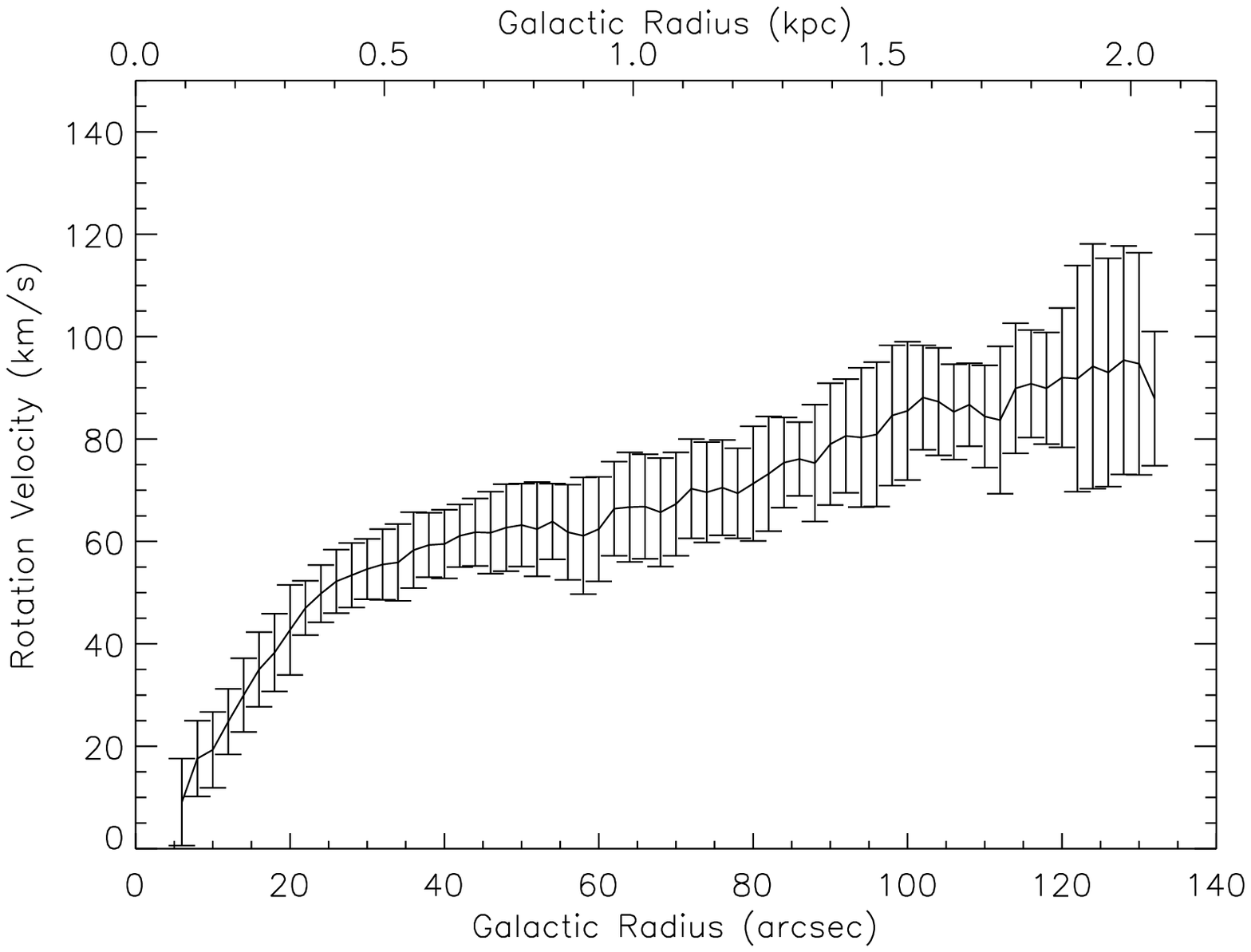}
} \vspace{1cm}

\caption{Rotation curves for the galaxies DDO 154 (top--left:
PA=236\Deg, i=59\Deg), NGC 4826 (top--right: PA=291\Deg, i=53\Deg),
NGC 5033 (bottom--left: PA=353\Deg, i=71\Deg), and NGC 7793
(bottom--right: PA=277\Deg, i=47\Deg).}
\end{figure*}

\label{lastpage}


\begin{thebibliography}{}

\bibitem[\protect\citeauthoryear{Afanasiev \& Sil'chenko}{1999}]{1999AJ....117.1725A} Afanasiev V.~L., Sil'chenko O.~K., 1999, AJ, 117, 1725
\bibitem[\protect\citeauthoryear{Afanasiev \& Sil'chenko}{2005}]{2005A&A...429..825A} Afanasiev V.~L., Sil'chenko O.~K., 2005, A\&A, 429, 825
\bibitem[\protect\citeauthoryear{Ag{\"u}ero, D{\'{\i}}az, \& Bajaja}{2004}]{2004A&A...414..453A} Ag{\"u}ero E.~L., D{\'{\i}}az R.~J., Bajaja E., 2004, A\&A, 414, 453
\bibitem[\protect\citeauthoryear{Aguerri, Beckman, \& Prieto}{1998}]{1998AJ....116.2136A} Aguerri J.~A.~L., Beckman J.~E., Prieto M., 1998, AJ, 116, 2136
\bibitem[\protect\citeauthoryear{Aguerri et al.}{2000}]{2000A&A...361..841A} Aguerri J.~A.~L., Mu{\~n}oz-Tu{\~n}{\'o}n C., Varela A.~M., Prieto M., 2000, A\&A, 361, 841
\bibitem[\protect\citeauthoryear{Barbieri et al.}{2005}]{2005A&A...439..947B} Barbieri C.~V., Fraternali F., Oosterloo T., Bertin G., Boomsma R., Sancisi R., 2005, A\&A, 439, 947
\bibitem[\protect\citeauthoryear{Blais--Ouellette, Amram, \& Carignan}{2001}]{2001AJ....121.1952B} Blais--Ouellette S., Amram P., Carignan C., 2001, AJ, 121, 1952
\bibitem[\protect\citeauthoryear{Bohuski et al.}{1972}]{1972ApJ...175..329B} Bohuski T.~J., Burbidge E.~M., Burbidge G.~R., Smith M.~G., 1972, ApJ, 175, 329
\bibitem[\protect\citeauthoryear{Boissier et al.}{2003}]{2003MNRAS.346.1215B} Boissier S., Prantzos N., Boselli A., Gavazzi G., 2003, MNRAS, 346, 1215
\bibitem[\protect\citeauthoryear{Boselli et al.}{2001}]{2001AJ....121..753B} Boselli A., Gavazzi G., Donas J., Scodeggio M. 2001, AJ, 121, 753
\bibitem[\protect\citeauthoryear{Bosma}{1981}]{1981AJ.....86.1791B} Bosma A., 1981, AJ, 86, 1791
\bibitem[\protect\citeauthoryear{Braun, Walterbos, \& Kennicutt}{1992}]{1992Natur.360..442B} Braun R., Walterbos R.~A.~M., Kennicutt R.~C., Jr., 1992, Natur, 360, 442
\bibitem[\protect\citeauthoryear{Bridges et al.}{1997}]{1997MNRAS.284..376B} Bridges T.~J., Ashman K.~M., Zepf S.~E., Carter D., Hanes D.~A., Sharples R.~M., Kavelaars J.~J., 1997, MNRAS, 284, 376
\bibitem[\protect\citeauthoryear{Bureau \& Carignan}{2002}]{2002AJ....123.1316B} Bureau M., Carignan C., 2002, AJ, 123, 1316
\bibitem[\protect\citeauthoryear{Burstein, Krumm, \& Salpeter}{1987}]{1987AJ.....94..883B} Burstein D., Krumm N., Salpeter E.~E., 1987, AJ, 94, 883
\bibitem[\protect\citeauthoryear{Buta}{1988}]{1988ApJS...66..233B} Buta R., 1988, ApJS, 66, 233
\bibitem[\protect\citeauthoryear{Carignan \& Beaulieu}{1989}]{1989ApJ...347..760C} Carignan C., Beaulieu S., 1989, ApJ, 347, 760
\bibitem[\protect\citeauthoryear{Carignan \& Freeman}{1988}]{1988ApJ...332L..33C} Carignan C., Freeman K.~C., 1988, ApJ, 332, L33
\bibitem[\protect\citeauthoryear{Carignan \& Puche}{1990}]{1990AJ....100..394C} Carignan C., Puche D., 1990, AJ, 100, 394
\bibitem[\protect\citeauthoryear{Carignan \& Purton}{1998}]{1998ApJ...506..125C} Carignan C., Purton C., 1998, ApJ, 506, 125
\bibitem[\protect\citeauthoryear{Carignan et al.}{2007}]{2007arXiv0705.4093C} Carignan C., Hernandez O., Beckman J.~E., Fathi K., 2007, arXiv, 705, arXiv:0705.4093
\bibitem[\protect\citeauthoryear{Carter \& Jenkins}{1993}]{1993MNRAS.263.1049C} Carter D., Jenkins C.~R., 1993, MNRAS, 263, 1049
\bibitem[\protect\citeauthoryear{Cayatte et al.}{1990}]{1990AJ....100..604C} Cayatte V., van Gorkom J.~H., Balkowski C., Kotanyi C., 1990, AJ, 100, 604
\bibitem[\protect\citeauthoryear{Chemin et al.}{2003}]{2003A&A...405...89C} Chemin L., Cayatte V., Balkowski C., Marcelin M., Amram P., van Driel W., Flores H., 2003, A\&A, 405, 89
\bibitem[\protect\citeauthoryear{Chemin et al.}{2006a}]{2006MNRAS.366..812C} Chemin L., et al., 2006a, MNRAS, 366, 812
\bibitem[\protect\citeauthoryear{Chemin et al.}{2006b}]{2006AJ....132.2527C} Chemin L., Carignan C., Drouin N., Freeman K.~C., 2006b, AJ, 132, 2527
\bibitem[\protect\citeauthoryear{Haynes, Giovanelli, \& Kent}{2007}]{2007ApJ...665L..19H} Haynes M.~P., Giovanelli R., Kent B.~R., 2007, ApJ, 665, L19
\bibitem[\protect\citeauthoryear{Colina et al.}{1997}]{1997ApJ...484L..41C} Colina L., Garcia Vargas M.~L., Mas--Hesse J.~M., Alberdi A., Krabbe A., 1997, ApJ, 484, L41
\bibitem[\protect\citeauthoryear{Daigle et al.}{2004}]{2004SPIE.5499..219D} Daigle O., Gach J.--L., Guillaume C., Carignan C., Balard P., Boisin O., 2004, SPIE, 5499, 219
\bibitem[\protect\citeauthoryear{Daigle et al.}{2006a}]{2006MNRAS.367..469D} Daigle O., Carignan C., Amram P., Hernandez O., Chemin L., Balkowski C., Kennicutt R., 2006a, MNRAS, 367, 469
\bibitem[\protect\citeauthoryear{Daigle et al.}{2006b}]{2006MNRAS.368.1016D} Daigle O., Carignan C., Hernandez O., Chemin L., Amram P., 2006b, MNRAS, 368, 1016
\bibitem[\protect\citeauthoryear{Daigle, Carignan, \& Blais--Ouellette}{2006}]{2006SPIE.6276E..42D} Daigle O., Carignan C., Blais--Ouellette S., 2006, SPIE, 6276,62761F
\bibitem[\protect\citeauthoryear{de Blok \& Walter}{2000}]{2000ApJ...537L..95D} de Blok W.~J.~G., Walter F., 2000, ApJ, 537, L95
\bibitem[\protect\citeauthoryear{Devereux, Kenney, \& Young}{1992}]{1992AJ....103..784D} Devereux N.~A., Kenney J.~D., Young J.~S., 1992, AJ, 103, 784
\bibitem[\protect\citeauthoryear{Dumas et al.}{2007}]{2007arXiv0705.4162D} Dumas G., Mundell C., Emsellem E., Nagar N., 2007, arXiv, 705, arXiv:0705.4162
\bibitem[\protect\citeauthoryear{Elfhag et al.}{1996}]{1996A&AS..115..439E} Elfhag T., Booth R.~S., Hoeglund B., Johansson L.~E.~B., Sandqvist A., 1996, A\&AS, 115, 439
\bibitem[\protect\citeauthoryear{Fathi et al.}{2006}]{2006ApJ...641L..25F} Fathi K., Storchi--Bergmann T., Riffel R.~A., Winge C., Axon D.~J., Robinson A., Capetti A., Marconi A., 2006, ApJ, 641, L25
\bibitem[\protect\citeauthoryear{Fathi et al.}{2007}]{2007A&A...466..905F} Fathi K., Beckman, J.~E., Zurita, A., Relano, M., Knapen, J.~H., Daigle, O., Hernandez, O., Carignan, C., 2007, A\&A, 466, 905
\bibitem[\protect\citeauthoryear{Gach et al.}{2002}]{2002PASP..114.1043G} Gach J.--L., et al., 2002, PASP, 114, 1043
\bibitem[\protect\citeauthoryear{Gerin, Combes, \& Nakai}{1988}]{1988A&A...203...44G} Gerin M., Combes F., Nakai N., 1988, A\&A, 203, 44
\bibitem[\protect\citeauthoryear{Golla, Dettmar, \& Domgoergen}{1996}]{1996A&A...313..439G} Golla G., Dettmar R.--J., Domgoergen H., 1996, A\&A, 313, 439
\bibitem[\protect\citeauthoryear{Gooch}{1996}]{1996ASPC..101...80G} Gooch R.~E., 1996, ASPC, 101, 80
\bibitem[\protect\citeauthoryear{Gordon}{1991}]{1991ApJ...371..563G} Gordon M.~A., 1991, ApJ, 371, 563
\bibitem[\protect\citeauthoryear{H{\"a}gele et al.}{2007}]{2007MNRAS.378..163H} H{\"a}gele G.~F., D{\'{\i}}az {\'A}.~I., Cardaci M.~V., Terlevich E., Terlevich R., 2007, MNRAS, 378, 163
\bibitem[\protect\citeauthoryear{Hawarden et al.}{1979}]{1979A&A....76..230H} Hawarden T.~G., van Woerden H., Goss W.~M., Mebold U., Peterson B.~A., 1979, A\&A, 76, 230
\bibitem[\protect\citeauthoryear{Haynes et al.}{2007}]{2007ApJ...665L..19H} Haynes M.~P., Giovanelli R., Kent B.~R. 2007, ApJ, 665, 19
\bibitem[\protect\citeauthoryear{Helfer et al.}{2003}]{2003ApJS..145..259H} Helfer T.~T., Thornley M.~D., Regan M.~W., Wong T., Sheth K., Vogel S.~N., Blitz L., Bock D.~C.--J., 2003, ApJS, 145, 259
\bibitem[\protect\citeauthoryear{Hernandez et al.}{2003}]{2003SPIE.4841.1472H} Hernandez O., Gach J.--L., Carignan C., Boulesteix J., 2003, SPIE, 4841, 1472
\bibitem[\protect\citeauthoryear{Hernandez et al.}{2005a}]{2005MNRAS.360.1201H} Hernandez O., Carignan C., Amram P., Chemin L., Daigle O., 2005a, MNRAS, 360, 1201
\bibitem[\protect\citeauthoryear{Hernandez et al.}{2005b}]{2005ApJ...632..253H} Hernandez O., Wozniak, H., Carignan C., Amram P., Chemin L., Daigle O., 2005b, ApJ, 632, 253
\bibitem[\protect\citeauthoryear{Hoffman et al.}{1993}]{1993AJ....106...39H} Hoffman G.~L., Lu N.~Y., Salpeter E.~E., Farhat B., Lamphier C., Roos T., 1993, AJ, 106, 39
\bibitem[\protect\citeauthoryear{Hota \& Saikia}{2005}]{2005MNRAS.356..998H} Hota A., Saikia D.~J., 2005, MNRAS, 356, 998
\bibitem[\protect\citeauthoryear{Jeong et al.}{2007}]{2007MNRAS.376.1021J} Jeong H., Bureau M., Yi S.~K., Krajnovi{\'c} D., Davies R.~L., 2007, MNRAS, 376, 1021
\bibitem[\protect\citeauthoryear{Kennicutt}{1989}]{1989ApJ...344..685K} Kennicutt R.~C., Jr., 1989, ApJ, 344, 685
\bibitem[\protect\citeauthoryear{Kennicutt}{1998a}]{1998ARA&A..36..189K} Kennicutt R.~C., Jr., 1998a, ARA\&A, 36, 189
\bibitem[\protect\citeauthoryear{Kennicutt}{1998b}]{1998ApJ...498..541K} Kennicutt R.~C., Jr., 1998b, ApJ, 498, 541
\bibitem[\protect\citeauthoryear{Kennicutt et al.}{2003}]{2003PASP..115..928K} Kennicutt R.~C., Jr., et al., 2003, PASP, 115, 928
\bibitem[\protect\citeauthoryear{Knapp}{1987}]{1987IAUS..127..145K} Knapp G.~R., 1987, IAUS, 127, 145
\bibitem[\protect\citeauthoryear{Kranz, Slyz, \& Rix}{2001}]{2001ApJ...562..164K} Kranz T., Slyz A., Rix H.--W., 2001, ApJ, 562, 164
\bibitem[\protect\citeauthoryear{Lake, Schommer, \& van Gorkom}{1987}]{1987ApJ...314...57L} Lake G., Schommer R.~A., van Gorkom J.~H., 1987, ApJ, 314, 57
\bibitem[\protect\citeauthoryear{Legrand et al.}{1997}]{1997A&A...326..929L} Legrand F., Kunth D., Mas--Hesse J.~M., Lequeux J., 1997, A\&A, 326, 929
\bibitem[\protect\citeauthoryear{Lequeux et al.}{1995}]{1995A&A...301...18L} Lequeux J., Kunth D., Mas--Hesse J.~M., Sargent W.~L.~W., 1995, A\&A, 301, 18
\bibitem[\protect\citeauthoryear{Leroy et al.}{2005}]{2005ApJ...625..763L} Leroy A., Bolatto A.~D., Simon J.~D., Blitz L., 2005, ApJ, 625, 763
\bibitem[\protect\citeauthoryear{Lindblad \& Jorsater}{1981}]{1981A&A....97...56L} Lindblad P.~O., Jorsater S., 1981, A\&A, 97, 56
\bibitem[\protect\citeauthoryear{Martimbeau, Carignan, \& Roy}{1994}]{1994AJ....107..543M} Martimbeau N., Carignan C., Roy J.--R., 1994, AJ, 107, 543
\bibitem[\protect\citeauthoryear{Martin}{1995}]{1995AJ....109.2428M} Martin P., 1995, AJ, 109, 2428
\bibitem[\protect\citeauthoryear{Martin \& Kennicutt}{2001}]{2001ApJ...555..301M} Martin C.~L., Kennicutt R.~C., 2001, ApJ, 555, 301
\bibitem[\protect\citeauthoryear{Mediavilla et al.}{2005}]{2005A&A...433...79M} Mediavilla E., Guijarro A., Castillo--Morales A., Jim{\'e}nez--Vicente J., Florido E., Arribas S., Garc{\'{\i}}a--Lorenzo B., Battaner E., 2005, A\&A, 433, 79
\bibitem[\protect\citeauthoryear{M{\'e}ndez \& Esteban}{2000}]{2000A&A...359..493M} M{\'e}ndez D.~I., Esteban C., 2000, A\&A, 359, 493
\bibitem[\protect\citeauthoryear{Men{\'e}ndez--Delmestre et al.}{2007}]{2007ApJ...657..790M} Men{\'e}ndez--Delmestre K., Sheth K., Schinnerer E., Jarrett T.~H., Scoville N.~Z., 2007, ApJ, 657, 790
\bibitem[\protect\citeauthoryear{Meurer, Staveley--Smith, \& Killeen}{1998}]{1998MNRAS.300..705M} Meurer G.~R., Staveley--Smith L., Killeen N.~E.~B., 1998, MNRAS, 300, 705
\bibitem[\protect\citeauthoryear{Meurer et al.}{2006}]{2006ApJS..165..307M} Meurer G.~R., et al., 2006, ApJS, 165, 307
\bibitem[\protect\citeauthoryear{Miller}{1995}]{1995ApJ...446L..75M} Miller B.~W., 1995, ApJ, 446, L75
\bibitem[\protect\citeauthoryear{Mu{\~n}oz--Tu{\~n}{\'o}n, Caon, \& Aguerri}{2004}]{2004AJ....127...58M} Mu{\~n}oz--Tu{\~n}{\'o}n C., Caon N., Aguerri J.~A.~L., 2004, AJ, 127, 58
\bibitem[\protect\citeauthoryear{Nakanishi et al.}{2007}]{2007PASJ...59...61N} Nakanishi H., Tosaki T., Kohno K., Sofue Y., Kuno N., 2007, PASJ, 59, 61
\bibitem[\protect\citeauthoryear{Ondrechen, van der Hulst, \& Hummel}{1989}]{1989ApJ...342...39O} Ondrechen M.~P., van der Hulst J.~M., Hummel E., 1989, ApJ, 342, 39
\bibitem[\protect\citeauthoryear{Palou{\v s}, Ehlerova, \& Elmegreen}{2002}]{2002Ap&SS.281..101P} Palou{\v s} J., Ehlerova S., Elmegreen B.~G., 2002, Ap\&SS, 281, 101
\bibitem[\protect\citeauthoryear{Pence, Taylor, \& Atherton}{1990}]{1990ApJ...357..415P} Pence W.~D., Taylor K., Atherton P., 1990, ApJ, 357, 415
\bibitem[\protect\citeauthoryear{Phookun, Vogel, \& Mundy}{1993}]{1993ApJ...418..113P} Phookun B., Vogel S.~N., Mundy L.~G., 1993, ApJ, 418, 113
\bibitem[\protect\citeauthoryear{Puche \& Carignan}{1988}]{1988AJ.....95.1025P} Puche D., Carignan C., 1988, AJ, 95, 1025
\bibitem[\protect\citeauthoryear{Rand}{1994}]{1994A&A...285..833R} Rand R.~J., 1994, A\&A, 285, 833
\bibitem[\protect\citeauthoryear{Regan et al.}{2002}]{2002ApJ...574..126R} Regan M.~W., Sheth K., Teuben P.~J., Vogel S.~N., 2002, ApJ, 574, 126
\bibitem[\protect\citeauthoryear{Regan et al.}{2006}]{2006ApJ...652.1112R} Regan M.~W., et al., 2006, ApJ, 652, 1112
\bibitem[\protect\citeauthoryear{Rownd, Dickey, \& Helou}{1994}]{1994AJ....108.1638R} Rownd B.~K., Dickey J.~M., Helou G., 1994, AJ, 108, 1638
\bibitem[\protect\citeauthoryear{Rubin}{1994}]{1994AJ....107..173R} Rubin V.~C., 1994, AJ, 107, 173
\bibitem[\protect\citeauthoryear{Rubin, Waterman, \& Kenney}{1999}]{1999AJ....118..236R} Rubin V.~C., Waterman A.~H., Kenney J.~D.~P., 1999, AJ, 118, 236
\bibitem[\protect\citeauthoryear{Sanders, Scoville, \& Soifer}{1991}]{1991ApJ...370..158S} Sanders D.~B., Scoville N.~Z., Soifer B.~T., 1991, ApJ, 370, 158
\bibitem[\protect\citeauthoryear{Shopbell \& Bland--Hawthorn}{1998}]{1998ApJ...493..129S} Shopbell P.~L., Bland--Hawthorn J., 1998, ApJ, 493, 129
\bibitem[\protect\citeauthoryear{Sofue et al.}{1990}]{1990PASJ...42..745S} Sofue Y., Handa T., Golla G., Wielebinski R., 1990, PASJ, 42, 745
\bibitem[\protect\citeauthoryear{Summers, Stevens, \& Strickland}{2001}]{2001astro.ph..6475S} Summers L.~K., Stevens I.~R., Strickland D.~K., 2001, astro, arXiv:astro-ph/0106475
\bibitem[\protect\citeauthoryear{Tempel \& Tenjes}{2006}]{2006MNRAS.371.1269T} Tempel E., Tenjes P., 2006, MNRAS, 371, 1269
\bibitem[\protect\citeauthoryear{Thean et al.}{1997}]{1997MNRAS.290...15T} Thean A.~H.~C., Mundell C.~G., Pedlar A., Nicholson R.~A., 1997, MNRAS, 290, 15
\bibitem[\protect\citeauthoryear{van der Hulst \& Huchtmeier}{1979}]{1979A&A....78...82V} van der Hulst J.~M., Huchtmeier W.~K., 1979, A\&A, 78, 82
\bibitem[\protect\citeauthoryear{van Driel \& Buta}{1991}]{1991A&A...245....7V} van Driel W., Buta R.~J., 1991, A\&A, 245, 7
\bibitem[\protect\citeauthoryear{van Driel, Rots, \& van Woerden}{1988}]{1988A&A...204...39V} van Driel W., Rots A.~H., van Woerden H., 1988, A\&A, 204, 39
\bibitem[\protect\citeauthoryear{Veilleux \& Rupke}{2002}]{2002ApJ...565L..63V} Veilleux S., Rupke D.~S., 2002, ApJ, 565, L63
\bibitem[\protect\citeauthoryear{Vogelaar \& Terlouw}{2001}]{2001ASPC..238..358V} Vogelaar M.~G.~R., Terlouw J.~P., 2001, ASPC, 238, 358
\bibitem[\protect\citeauthoryear{Walsh et al.}{1990}]{1990ApJ...352..532W} Walsh D.~E.~P., van Gorkom J.~H., Bies W.~E., Katz N., Knapp G.~R., Wallington S., 1990, ApJ, 352, 532
\bibitem[\protect\citeauthoryear{Walter, Weiss, \& Scoville}{2002}]{2002ApJ...580L..21W} Walter F., Weiss A., Scoville N., 2002, ApJ, 580, L21
\bibitem[\protect\citeauthoryear{Walter et al.}{2007}]{2007ApJ...661..102W} Walter F., et al., 2007, ApJ, 661, 102
\bibitem[\protect\citeauthoryear{Wong \& Blitz}{2000}]{2000ApJ...540..771W} Wong T., Blitz L., 2000, ApJ, 540, 771
\bibitem[\protect\citeauthoryear{Zhang, Wright, \& Alexander}{1993}]{1993ApJ...418..100Z} Zhang X., Wright M., Alexander P., 1993, ApJ, 418, 100

\end{thebibliography}
\end{document}